\documentclass[preprint,showpacs,titlepage,aps,prd,tightenlines,amsmath,byrevtex,nofootinbib]{revtex4}
\usepackage{graphicx}

\begin{document}

\preprint{hep-ph/0405005}

\title{
New plots and
parameter degeneracies in neutrino oscillations
}

\author{Osamu~Yasuda}
\email{E-mail: yasuda@phys.metro-u.ac.jp}
\affiliation{Department of Physics, Tokyo Metropolitan University,
Hachioji, Tokyo 192-0397, Japan}

\date{\today}

\vglue 1.4cm
\begin{abstract}
It is shown that eightfold degeneracy in neutrino oscillations
is easily seen by plotting constant probabilities in the
$(\sin^22\theta_{13}, 1/s^2_{23})$ plane. Using this plot,
we discuss how an additional long baseline
measurement resolves degeneracies after the JPARC
experiment measures the oscillation probabilities
$P(\nu_\mu\rightarrow\nu_e)$ and
$P(\bar{\nu}_\mu\rightarrow\bar{\nu}_e)$ at $|\Delta m^2_{31}|L/4E=\pi/2$.
By measuring
$P(\nu_\mu\rightarrow\nu_e)$ or $P(\bar{\nu}_\mu\rightarrow\bar{\nu}_e)$,
the sgn($\Delta m_{31}^2$) ambiguity is resolved better at
longer baselines and the $\delta\leftrightarrow\pi-\delta$
ambiguity is resolved better when $\left||\Delta m^2_{31}|L/4E-\pi/2\right|$
is larger.  The $\theta_{23}$ ambiguity may be resolved as a byproduct
if $\left||\Delta m^2_{31}|L/4E-\pi\right|$ is small and
the CP phase $\delta$ turns out to satisfy
$\left|\cos(\delta +|\Delta m^2_{31}|L/4E)\right|\sim1$.
It is pointed out that the low energy option ($E\sim$1GeV) at the off-axis
NuMI experiment may be useful in resolving these ambiguities.
The $\nu_e\rightarrow\nu_\tau$ channel offers a promising
possibility which may potentially resolve all the ambiguities.
\end{abstract}


\maketitle


\section{Introduction}

\begin{figure}
\includegraphics[scale=0.7]{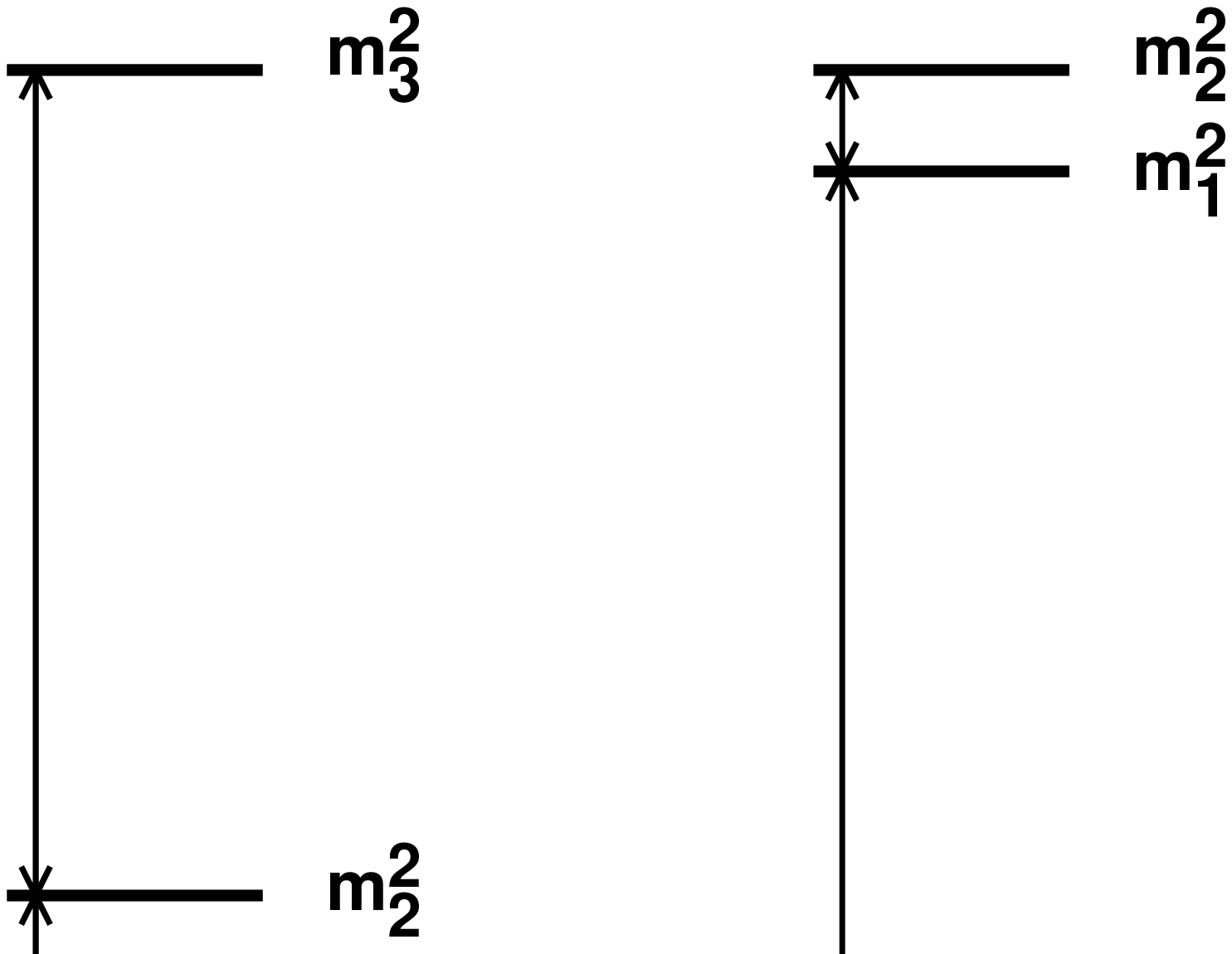}
\vglue 4.5cm
\caption{Two mass patterns.  
(a), (b) correspond to the normal ($\Delta m^2_{31}>0$), 
inverted ($\Delta m^2_{31}<0$) hierarchy,
respectively.
}
\label{fig1}
\end{figure}

From the recent experiments on atmospheric \cite{Kajita:2001mr} and
solar \cite{Bahcall:2000kh}, and
reactor \cite{Apollonio:1999ae,Eguchi:2002dm} neutrinos,
we now know approximately
the values of the mixing angles and the mass squared differences
of the atmospheric and solar neutrino oscillations:

\begin{eqnarray}
(\sin^22\theta_{12}, \Delta m^2_{21})&\simeq&
(0.8, 7\times10^{-5}{\rm eV}^2) ~\qquad\mbox{\rm for the solar neutrino}
\nonumber\\
(\sin^22\theta_{23}, |\Delta m^2_{31}|)&\simeq&
(1.0, 2\times10^{-3}{\rm eV}^2) ~\qquad\mbox{\rm for the atmospheric neutrino},
\nonumber
\end{eqnarray}
where
we use the standard parametrization \cite{Hagiwara:pw}
of the MNS mixing matrix
\begin{eqnarray}
U=\left(
\begin{array}{ccc}
c_{12}c_{13} & s_{12}c_{13} &  s_{13}e^{-i\delta}\\
-s_{12}c_{23}-c_{12}s_{23}s_{13}e^{i\delta} & 
c_{12}c_{23}-s_{12}s_{23}s_{13}e^{i\delta} & s_{23}c_{13}\\
s_{12}s_{23}-c_{12}c_{23}s_{13}e^{i\delta} & 
-c_{12}s_{23}-s_{12}c_{23}s_{13}e^{i\delta} & c_{23}c_{13}
\end{array}\right),
\nonumber
\end{eqnarray}
and the case of $\Delta m_{31}^2>0$ ($\Delta m_{31}^2<0$)
corresponds to the normal (inverted) mass hierarchy, as is
shown in Fig.\,\ref{fig1}.
In the three flavor framework of neutrino oscillations,
the oscillation parameters which are still unknown to date are
the third mixing angle $\theta_{13}$, the sign
of the mass squared difference $\Delta m^2_{31}$ 
of the atmospheric neutrino oscillation, and
the CP phase $\delta$.
It is expected that long baseline experiments in the
future will determine these three quantities.

Since the work of \cite{Burguet-Castell:2001ez}, it has been
known that even if the values of the oscillation
probabilities $P(\nu_\mu \rightarrow \nu_e)$ and
$P(\bar{\nu}_\mu \rightarrow \bar{\nu}_e)$ are exactly given
we cannot determine uniquely the values of the
oscillation parameters due to parameter degeneracies.
There are three kinds of parameter degeneracies:
the intrinsic $(\theta_{13}, \delta)$ 
degeneracy \cite{Burguet-Castell:2001ez},
the degeneracy of
$\Delta m^2_{31}\leftrightarrow-\Delta m^2_{31}$ \cite{Minakata:2001qm},
and the degeneracy of
$\theta_{23}\leftrightarrow\pi/2
-\theta_{23}$ \cite{Fogli:1996pv,Barger:2001yr}.
The intrinsic degeneracy is exact when $\Delta m^2_{21}/\Delta m^2_{31}$
is exactly zero.
The sgn($\Delta m^2_{31}$) degeneracy is exact when
$AL$ is exactly zero, where $A\equiv\sqrt{2}G_FN_e$ and $L$ stand for
the matter effect and the baseline, respectively ($G_F$ is the Fermi
constant and $N_e$ is the electron density in matter).
The $\theta_{23}$ degeneracy is exact when $\cos2\theta_{23}$
is exactly zero.
Each degeneracy gives a twofold solution, so
in total we have an eightfold solution if
all the degeneracies are exact.  In this case
prediction for physics is the same for all the
degenerated solutions and there is no problem.
However, these degeneracies
are lifted slightly in long baseline experiments \footnote{
$\cos2\theta_{23}$ may be exactly zero, but the present atmospheric
neutrino data \cite{saji} allow the possibility of $\cos2\theta_{23}\ne0$, so
we will assume $\cos2\theta_{23}\ne0$ in general in the following
discussions.}, and there are in general eight
different solutions \cite{Barger:2001yr}.
When we try to determine the oscillation parameters,
ambiguities arise because the values of the oscillation parameters are
slightly different for each solution.
In particular, this causes a serious problem in measurement of
CP violation, which is expected to be small effect
in the long baseline experiments, and we could mistake
a fake effect due to the ambiguities for
nonvanishing CP violation if we do not treat the ambiguities carefully.

In the references \cite{Burguet-Castell:2001ez,Minakata:2001qm,Barger:2001yr} in the past,
various diagrams have been given to
visualize how degeneracies are lifted in the parameter space.
To see how the eightfold degeneracy is lifted, it is necessary for
the plot to give eight different points for different eight solutions.
An effort was made in \cite{Minakata:2002jv} to visualize the
eight different points by plotting the trajectories of
constant probabilities in the $(\sin^22\theta_{13}, s^2_{23})$ plane.
In the present paper we propose a plot in the
$(\sin^22\theta_{13}, 1/s^2_{23})$ plane, which offers the simplest
way to visualize how the eightfold degeneracy is lifted.
As a byproduct, we show how the third measurement of
$\nu_\mu\rightarrow\nu_e$, $\bar{\nu}_\mu\rightarrow\bar{\nu}_e$
or $\nu_e\rightarrow\nu_\tau$ resolves the ambiguities, after
the JPARC experiment \cite{Itow:2001ee} measures the oscillation probabilities
$P(\nu_\mu\rightarrow\nu_e)$ and
$P(\bar{\nu}_\mu\rightarrow\bar{\nu}_e)$ at the
oscillation maximum, i.e., at $|\Delta m^2_{31}|L/4E=\pi/2$.

In the following discussions we assume that
$|\Delta m^2_{31}|$, $\Delta m^2_{21}$ and $\theta_{12}$ are
sufficiently precisely known.  This is justified because
the correlation between these parameters and the CP phase
$\delta$ is not so strong in the case of JPARC \cite{Pinney:2001xw}, and
we can safely ignore the uncertainty of these
parameters to discuss the ambiguities in $\delta$
due to parameter degeneracies.

\section{Plots in the $(\sin^22\theta_{13}, 1/s^2_{23})$ plane}
As in Ref. \cite{Yasuda:2003qg}, let us discuss the ambiguities
due to degeneracies step by step in the order
$(\theta_{23}-\pi/4=0, \Delta m^2_{21}=0, A=0)$
$\rightarrow$
$(\theta_{23}-\pi/4\ne0, \Delta m^2_{21}=0, A=0)$
$\rightarrow$
$(\theta_{23}-\pi/4\ne0, \Delta m^2_{21}\ne0, A=0)$
$\rightarrow$
$(\theta_{23}-\pi/4\ne0, \Delta m^2_{21}\ne0, A\ne0)$.

\subsection{$\cos2\theta_{23}=0, \Delta m^2_{21}/\Delta m^2_{31}=0,
AL=0$}
\begin{figure}
\vglue 1cm
\includegraphics[scale=1.0]{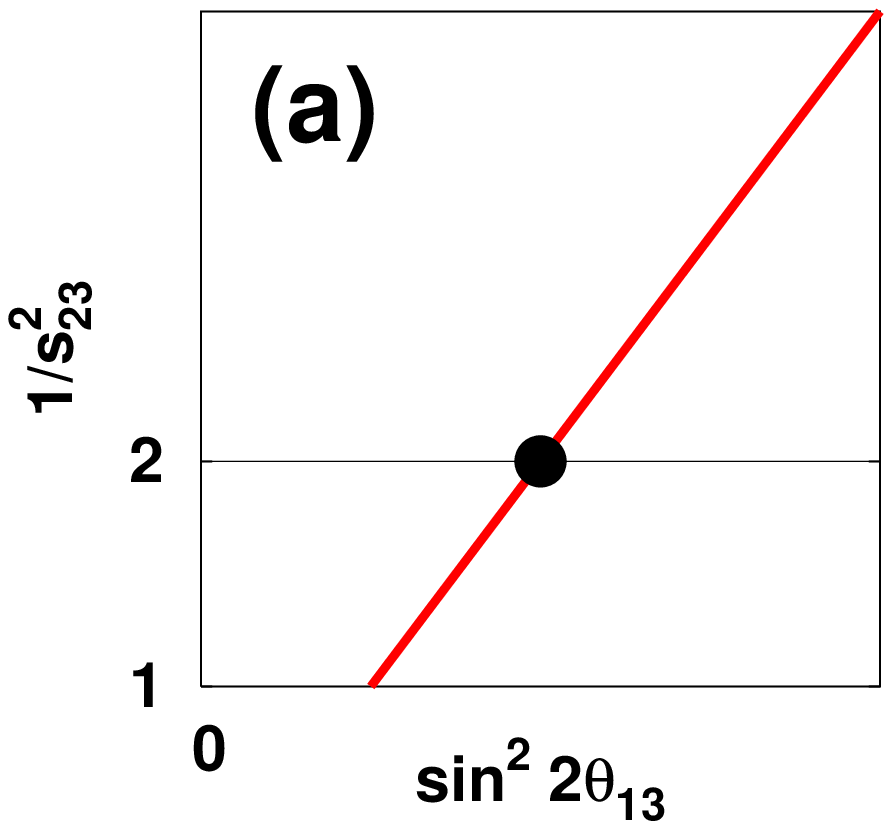}
\includegraphics[scale=1.0]{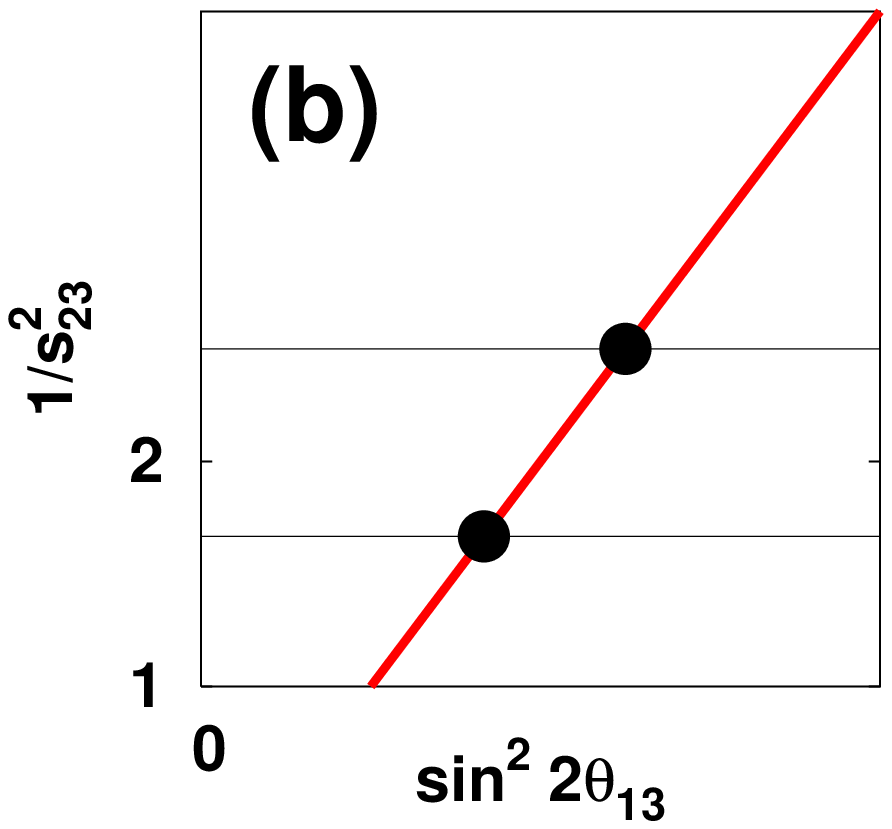}
\vglue 0.5cm
\caption{The solutions, which are marked by black blobs,
for given $P(\nu_\mu\rightarrow\nu_e)$,
$P(\bar{\nu}_\mu\rightarrow\bar{\nu}_e)$ and 
$P(\nu_\mu\rightarrow\nu_\mu)$ in the case
of $\Delta m^2_{21}/\Delta m^2_{31}=0, AL=0$.
(a) For $\cos2\theta_{23}=0$, the intersection of $Y\equiv 1/s^2_{23}=2$ and the
trajectory of
$P(\nu_\mu\rightarrow\nu_e)=P(\nu_\mu\rightarrow\nu_\mu)=$ const.
is one point with eightfold degeneracy.
(b) For $\cos2\theta_{23}\ne0$, the intersections
are two solutions with fourfold degeneracy.
}
\label{fig2}
\end{figure}

In this case the oscillation probabilities $P(\nu_\mu\rightarrow\nu_e)$
and $P(\bar{\nu}_\mu\rightarrow\bar{\nu}_e)$ are equal and are given by
\begin{eqnarray}
P(\nu_\mu\rightarrow\nu_e)=
P(\bar{\nu}_\mu\rightarrow\bar{\nu}_e)
=s_{23}^2\sin^22\theta_{13}\sin^2\Delta,\nonumber
\end{eqnarray}
where we have introduced the notation
\begin{eqnarray}
\Delta\equiv\frac{|\Delta m_{31}|L}{4E}.
\nonumber
\end{eqnarray}
To plot the line $P(\nu_\mu\rightarrow\nu_e)=
P(\bar{\nu}_\mu\rightarrow\bar{\nu}_e)$=const.
in the ($\sin^22\theta_{13}$, $1/s^2_{23}$) plane,
let us introduce the variables
\begin{eqnarray}
X&\equiv&\sin^22\theta_{13},\nonumber\\
Y&\equiv&\frac{1}{s^2_{23}}.
\nonumber
\end{eqnarray}
Then
\begin{eqnarray}
P=s_{23}^2\sin^22\theta_{13}\sin^2\Delta
\nonumber
\end{eqnarray}
give a straight line
\begin{eqnarray}
Y=\frac{\sin^2\Delta}{P}X
\label{degene1}
\end{eqnarray}
in the ($X$, $Y$) plane,
where $P$ and $\sin^2\Delta$ are constant.
The intersection of Eq. (\ref{degene1}) and
$Y\equiv1/s_{23}^2=2$ in the ($\sin^22\theta_{13}$, $1/s^2_{23}$) plane
is a unique point, which corresponds to a solution with
eightfold degeneracy.  The solution is depicted in Fig.\,\ref{fig2}(a).

\subsection{$\cos2\theta_{23}\ne0, \Delta m^2_{21}/\Delta m^2_{31}=0,
AL=0$}
At present the Superkamiokande atmospheric
neutrino data gives the allowed region
$0.90<\sin2\theta_{23}\le 1.0$ at 90\%CL \cite{saji}, and 
$\sin^22\theta_{23}$ can be in general different from 1.0.
If $\sin^22\theta_{23}$, which is more accurately
determined from the oscillation
probability $P(\nu_\mu\rightarrow\nu_\mu)$ in the future
long baseline experiments, deviates from 1, then we have
two solutions for $Y\equiv1/s^2_{23}$:
\begin{eqnarray}
Y_+&=&\frac{2}{1-\sqrt{1-\sin^22\theta_{23}}}
\nonumber\\
Y_-&=&\frac{2}{1+\sqrt{1-\sin^22\theta_{23}}}.
\nonumber
\end{eqnarray}
In this case there are two solutions, the one given by
Eq. (\ref{degene1}) and $Y=Y_+$ and another given by
Eq. (\ref{degene1}) and $Y=Y_-$.  These are two solutions
with fourfold degeneracy.  The two solutions in the
($\sin^22\theta_{13}$, $1/s^2_{23}$) plane are shown in
Fig.\,\ref{fig2}(b).  From this we see that even if we know
precisely the values of $P(\nu_\mu\rightarrow\nu_e)$,
$P(\bar{\nu}_\mu\rightarrow\bar{\nu}_e)$ and $P(\nu_\mu\rightarrow\nu_\mu)$,
there are two sets of solutions, and this is the ambiguity
due to the $\theta_{23}\leftrightarrow\pi-\theta_{23}$ degeneracy.

\subsection{$\cos2\theta_{23}\ne0, \Delta m^2_{21}/\Delta m^2_{31}\ne0,
AL=0$}
If we turn on the effect of non-zero $\Delta m^2_{21}$
in addition to non-zero $\cos2\theta_{23}$, then the oscillation
probabilities are \footnote{This is obtained by taking the limit
$A\equiv\sqrt{2}G_FN_e\rightarrow0$ in Eq. (16) in Ref. \cite{Cervera:2000kp}.}
\begin{eqnarray}
\left\{ \begin{array}{c}
P(\nu_\mu \to \nu_e)\\
P(\bar\nu_\mu \to \bar\nu_e)
\end{array}\right\}
=x^2 \sin^2\Delta + 2 x y \Delta
\sin\Delta\cos\left(\delta\pm\Delta\right)
+ y^2 \Delta^2\,,\nonumber
\end{eqnarray}
which are correct to the second order in the small parameters
$|\Delta m^2_{21}/\Delta m^2_{31}|$ and
$\sin 2\theta_{13}$,
where
\begin{eqnarray}
x&\equiv&s_{23} \sin 2\theta_{13} \,,
\nonumber\\
y &\equiv& \left|\frac{\Delta m_{21}^2}{\Delta m_{31}^2}\right|
c_{23} \sin 2\theta_{12}.
\label{xy}
\end{eqnarray}
In this case, the trajectory of $P(\nu_\mu \to \nu_e)=P$,
$P(\bar\nu_\mu \to \bar\nu_e)=\bar{P}$, where $P$ and $\bar{P}$
are constant, in the ($X\equiv\sin^22\theta_{13}$, $Y\equiv1/s^2_{23}$) plane
is given by a quadratic curve:
\begin{eqnarray}
&{\ }&16C_0X(Y-1)\Delta^2\sin^2\Delta\nonumber\\
&=&\frac{1}{\sin^2\Delta}(P-\bar{P})^2Y^2
+\frac{1}{\cos^2\Delta}\left[(P+\bar{P}-2C_0)(Y-1)+P+\bar{P}
-2X\sin^2\Delta\right]^2,
\label{degene2}
\end{eqnarray}
where
\begin{eqnarray}
C_0\equiv\left(\frac{\Delta m^2_{21}}{\Delta m^2_{31}}
\right)^2\Delta^2\sin^22\theta_{12}.
\nonumber
\end{eqnarray}
Eq. (\ref{degene2}) becomes a hyperbola for most of the range of $\Delta$,
but it becomes an ellipse for some region $\Delta\simeq\pi$.

\begin{figure}
\vglue 1cm
\includegraphics[scale=1.0]{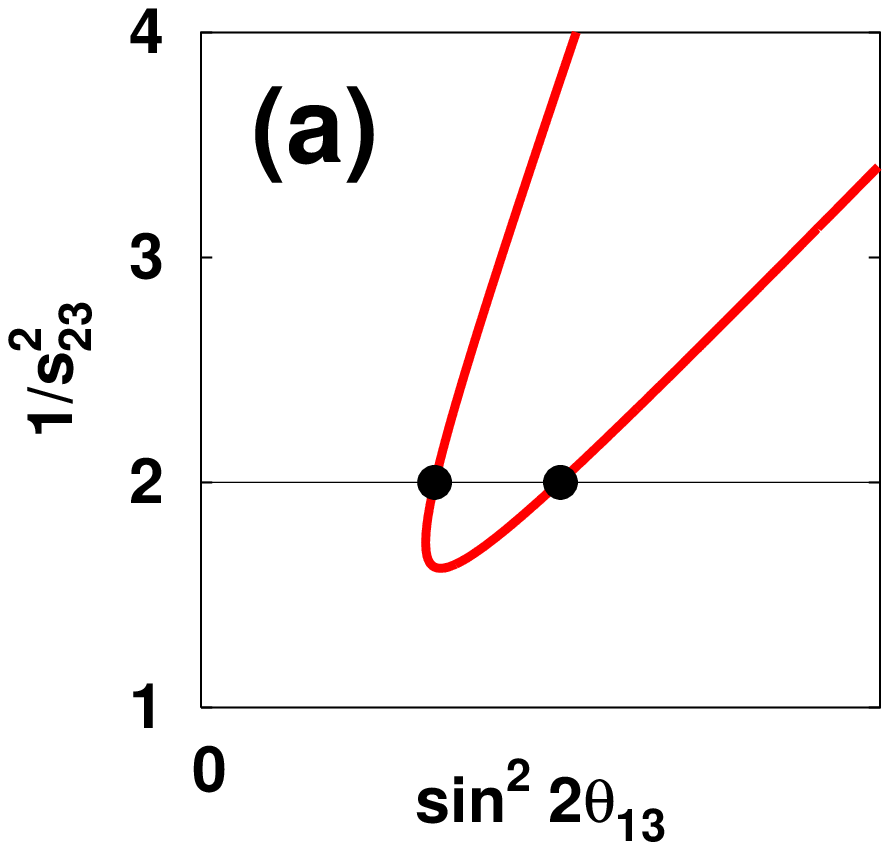}
\includegraphics[scale=1.0]{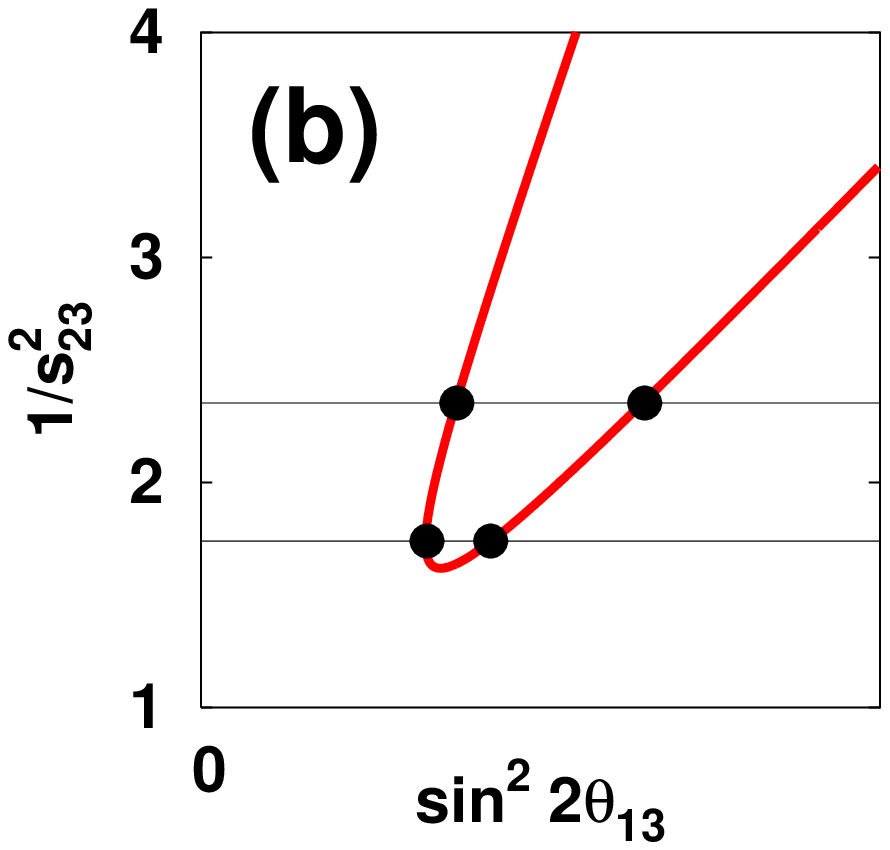}
\vglue 1.0cm
\caption{The solutions, which are marked by black blobs,
for given $P(\nu_\mu\rightarrow\nu_e)$,
$P(\bar{\nu}_\mu\rightarrow\bar{\nu}_e)$ and 
$P(\nu_\mu\rightarrow\nu_\mu)$ in the case
of $\Delta m^2_{21}/\Delta m^2_{31}\ne0, AL=0$.
(a) For $\cos2\theta_{23}=0$, the intersection of $Y\equiv 1/s^2_{23}=2$ and the
trajectory of
$P(\nu_\mu\rightarrow\nu_e)=$ const., 
$P(\nu_\mu\rightarrow\nu_\mu)=$ const.
are two points with fourfold degeneracy.
(b) For $\cos2\theta_{23}\ne0$, the intersections
are four solutions with twofold degeneracy.
}
\label{fig3}
\end{figure}

When $\sin^22\theta_{23}=1$, there are two solutions for the intersection
of $Y=2$ and Eq. (\ref{degene2}).  This indicates that even if we know
the precise values of $P(\nu_\mu\rightarrow\nu_e)$,
$P(\bar{\nu}_\mu\rightarrow\bar{\nu}_e)$ and $P(\nu_\mu\rightarrow\nu_\mu)$,
there are two sets of solutions for $(\theta_{13}, \theta_{23}, \delta)$
with fourfold degeneracy when $\sin^22\theta_{23}=1$,
as is depicted in Fig.\,\ref{fig3}(a).
This is the ambiguity
due to the intrinsic $(\theta_{13}, \delta)$ degeneracy.
When $\sin^22\theta_{23}\ne1$, there are four sets of solutions with
twofold degeneracy, as is depicted in Fig.\,\ref{fig3}(b).

\subsection{$\cos2\theta_{23}\ne0, \Delta m^2_{21}/\Delta m^2_{31}\ne0,
AL\ne0$}
Furthermore, if we turn on the matter effect $AL$,
then the oscillation
probabilities are given by \cite{Cervera:2000kp,Barger:2001yr}
\begin{eqnarray}
P(\nu_\mu \to \nu_e)
&=&x^2 f^2 + 2 x y f g
\cos\left(\delta+\Delta\right)
+ y^2 g^2\,,\nonumber\\
P(\bar\nu_\mu \to \bar\nu_e) &=&
x^2 \bar{f}^2+ 2 x y \bar{f}g  
\cos\left(\delta-\Delta\right)+ y^2 g^2 \,,
\label{degene3n}
\end{eqnarray}
for the normal hierarchy, while
\begin{eqnarray}
P(\nu_\mu \to \nu_e)
&=&x^2 \bar{f}^2- 2 x y \bar{f}g
\cos\left(\delta-\Delta\right)
+ y^2 g^2\,,\nonumber\\
P(\bar\nu_\mu \to \bar\nu_e) &=&
x^2 f^2 - 2 x y f g  
\cos\left(\delta+\Delta\right)+ y^2 g^2 \,,
\label{degene3i}
\end{eqnarray}
for the inverted hierarchy,
where $x$ and $y$ are
given by Eq. (\ref{xy}), and
\begin{eqnarray}
\left\{ \begin{array}{c}
f\\ \bar{f}\end{array}\right\}
&\equiv& \frac{\sin\left(\Delta\mp AL/2\right)}
{\left(1\mp AL/2\Delta\right)}\,,
\label{f}\\
g &\equiv& \frac{\sin\left(AL/2\right)}
{AL/2\Delta}\,.\label{g}
\end{eqnarray}
Eqs. (\ref{degene3n}) and (\ref{degene3i}) are correct
up to the second order in
$|\Delta m^2_{21}/\Delta m^2_{31}|$ and $\sin 2\theta_{13}$,
and all orders in $AL$.
The trajectory of $P(\nu_\mu \to \nu_e)=P$,
$P(\bar\nu_\mu \to \bar\nu_e)=\bar{P}$, where $P$ and $\bar{P}$
are constant, in the ($X\equiv\sin^22\theta_{13}$, $Y\equiv1/s^2_{23}$) plane
is again a quadratic curve for either of the mass hierarchies:
\begin{eqnarray}
\hspace*{-10mm}
16CX(Y-1)
&=&\frac{1}{\cos^2\Delta}\left[\left(\frac{P-C}{f}
+\frac{\bar{P}-C}{\bar{f}}\right)(Y-1)-(f+\bar{f})X
+\frac{P}{f}+\frac{\bar{P}}{\bar{f}}\right]^2\nonumber\\
&{\ }&+\frac{1}{\sin^2\Delta}\left[\left(\frac{P-C}{f}
-\frac{\bar{P}-C}{\bar{f}}\right)(Y-1)-(f-\bar{f})X
+\frac{P}{f}-\frac{\bar{P}}{\bar{f}}\right]^2
\label{degene4n}
\end{eqnarray}
for the normal hierarchy, and
\begin{eqnarray}
\hspace*{-10mm}
16CX(Y-1)
&=&\frac{1}{\cos^2\Delta}\left[\left(\frac{P-C}{\bar{f}}
+\frac{\bar{P}-C}{f}\right)(Y-1)-(f+\bar{f})X
+\frac{P}{\bar{f}}+\frac{\bar{P}}{f}\right]^2\nonumber\\
&{\ }&+\frac{1}{\sin^2\Delta}\left[\left(\frac{P-C}{\bar{f}}
-\frac{\bar{P}-C}{f}\right)(Y-1)-(f-\bar{f})X
+\frac{P}{\bar{f}}-\frac{\bar{P}}{f}\right]^2
\label{degene4i}
\end{eqnarray}
for the inverted hierarchy, where
\begin{eqnarray}
C&\equiv&\left(\frac{\Delta m^2_{21}}{\Delta m^2_{31}}
\right)^2\left[\frac{\sin(AL/2)}{AL/2\Delta}
\right]^2\sin^22\theta_{12}.
\label{c}
\end{eqnarray}
Again these quadratic curves become hyperbolas for most
of the region of $\Delta$, but they become ellipses for
some $\Delta\simeq\pi$.

\begin{figure}
\vglue 1cm
\includegraphics[scale=1.0]{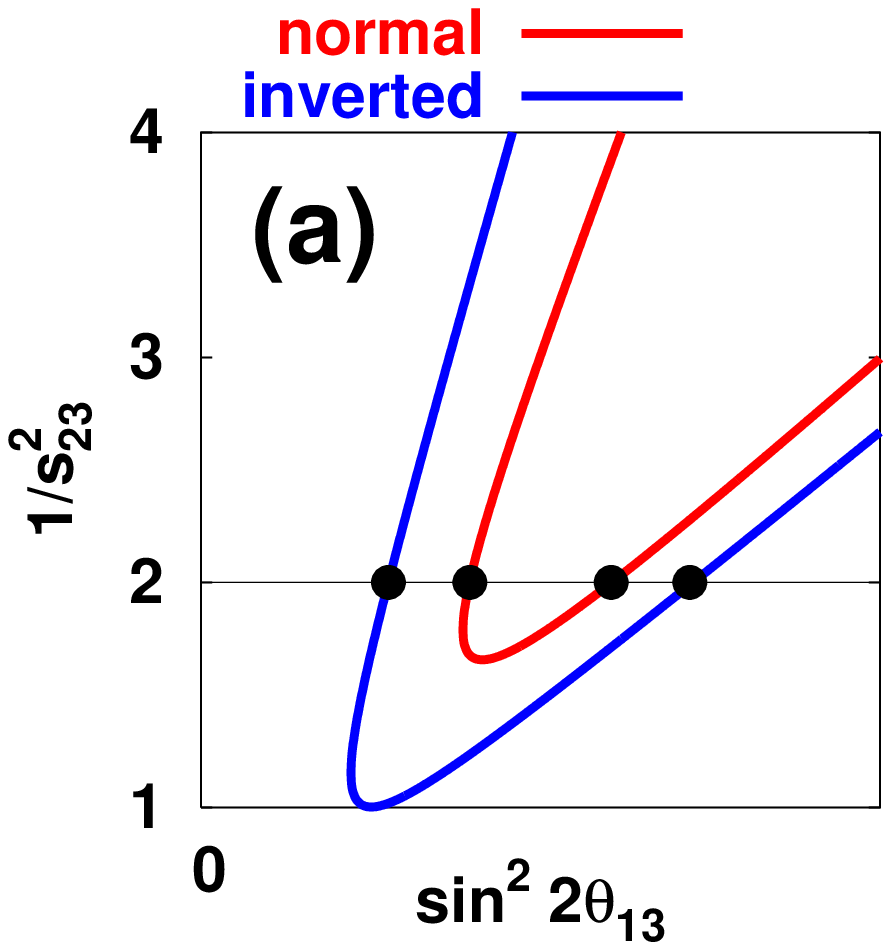}
\includegraphics[scale=1.0]{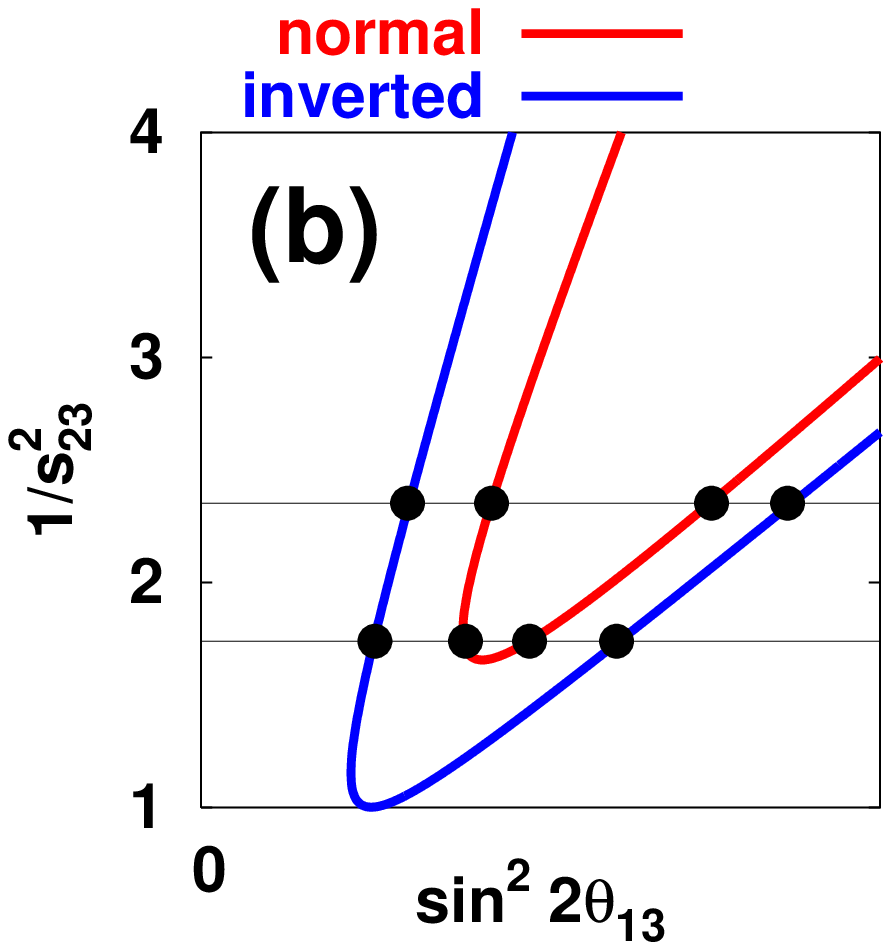}
\vglue 1.0cm
\caption{The solutions, which are marked by black blobs,
for given $P(\nu_\mu\rightarrow\nu_e)$,
$P(\bar{\nu}_\mu\rightarrow\bar{\nu}_e)$ and 
$P(\nu_\mu\rightarrow\nu_\mu)$ in the case
of $\Delta m^2_{21}/\Delta m^2_{31}\ne0, AL\ne0$.
(a) For $\cos2\theta_{23}=0$, the intersection of $Y\equiv 1/s^2_{23}=2$ and the
trajectory of
$P(\nu_\mu\rightarrow\nu_e)=$ const.,
$P(\nu_\mu\rightarrow\nu_\mu)=$ const.
four points with twofold degeneracy.
(b) For $\cos2\theta_{23}\ne0$, the intersections
are eight solutions without degeneracy.
}
\label{fig4}
\end{figure}

If $\sin^22\theta_{23}=1$, then there are four solutions with
twofold degeneracy, as is shown in Fig.\,\ref{fig4}(a).
If we know for some reason
(e.g., from reactor experiments) which solution is selected
for each mass hierarchy, there are only two solutions.  This
is the ambiguity due to the sgn($\Delta m^2_{31}$) degeneracy.
If $\sin^22\theta_{23}\ne1$ and if we do not know which solution
is favored with respect to the intrinsic degeneracy for each hierarchy,
and if we do not know sgn($\Delta m^2_{31}$),
then there are eight solutions without any degeneracy,
as is depicted in Fig.\,\ref{fig4}(b).
The advantage of our plot is that all the eight solutions
for $(\theta_{13}, \theta_{23})$ give different points,
and all the lines
in the ($\sin^22\theta_{13}$, $1/s^2_{23}$) plane
are described by (at most) quadratic curves
so that their behaviors are easy to see.

\subsection{Oscillation maximum\label{om}}
Finally, let us consider the case where experiments are
done at the oscillation maximum, i.e., when the neutrino
energy$E$ satisfies $\Delta\equiv|\Delta m^2_{31}|L/4E=\pi/2$.
In this case, the probabilities become
\begin{eqnarray}
P(\nu_\mu \to \nu_e)
&=&x^2 f^2 - 2 x y f g
\sin\delta
+ y^2 g^2\,,\label{om1}\\
P(\bar\nu_\mu \to \bar\nu_e) &=&
x^2 \bar{f}^2+ 2 x y \bar{f}g  
\sin\delta+ y^2 g^2 \,,
\label{om2}
\end{eqnarray}
for the normal hierarchy, and
\begin{eqnarray}
P(\nu_\mu \to \nu_e)
&=&x^2 \bar{f}^2- 2 x y \bar{f}g
\sin\delta
+ y^2 g^2\,,\label{om3}\\
P(\bar\nu_\mu \to \bar\nu_e) &=&
x^2 f^2 + 2 x y f g  
\sin\delta+ y^2 g^2 \,,
\label{om4}
\end{eqnarray}
for the inverted hierarchy, where $x$ and $y$ are
given by Eq. (\ref{xy}),
and $f$, $\bar{f}$, $g$ in Eqs. (\ref{f}), (\ref{g})
become
\begin{eqnarray}
\left\{ \begin{array}{c}
f\\ \bar{f} \end{array}\right\}
= \pm\frac{\cos(AL/2)}{1\mp AL/\pi}\,,~~
g &\equiv& \frac{\sin\left(AL/2\right)}
{AL/\pi}
\nonumber
\end{eqnarray}
for $\Delta=\pi/2$.
The trajectory of $P(\nu_\mu \to \nu_e)=P$,
$P(\bar\nu_\mu \to \bar\nu_e)=\bar{P}$
in the ($X\equiv\sin^22\theta_{13}$, $Y\equiv1/s^2_{23}$) plane
becomes a straight line and is given by
\begin{eqnarray}
Y=\frac{f+\bar{f}}{P/f+\bar{P}/\bar{f}-C(1/f+1/\bar{f})}
\left(X-\frac{C}{f\bar{f}}\right)
\label{omn}
\end{eqnarray}
for the normal hierarchy, and
\begin{eqnarray}
Y=\frac{f+\bar{f}}{P/\bar{f}+\bar{P}/f-C(1/f+1/\bar{f})}
\left(X-\frac{C}{f\bar{f}}\right)
\label{omi}
\end{eqnarray}
for the inverted hierarchy, where $C$ is given in Eq. (\ref{c}).
The straight lines (\ref{omn}) and (\ref{omi}) are extremely
close to each other in relatively short long baseline experiments
such as JPARC, where the matter effect is small.
As is shown in Appendix \ref{appendix2}, (\ref{omn}) and (\ref{omi})
have the minimum values in $Y\equiv1/s^2_{23}$ which is
larger than the naive value 1 for either of the mass hierarchies.
Since Eqs. (\ref{omn}) and (\ref{omi}) are linear in $X$,
there is only one solution between them and $Y$=const.
Thus the ambiguity due to the intrinsic degeneracy is solved
by performing experiments at the oscillation maximum, although
it is then transformed into
another ambiguity due to the $\delta\leftrightarrow\pi-\delta$ degeneracy.

\begin{figure}
\includegraphics[scale=1.0]{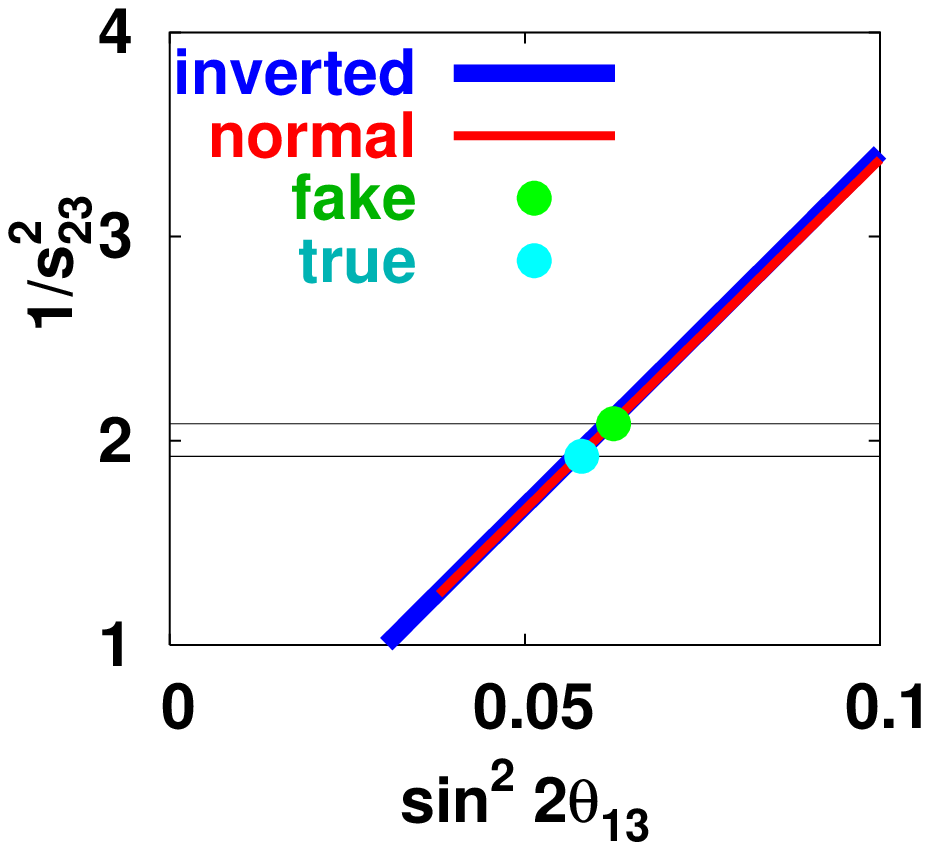}
\includegraphics[scale=1.0]{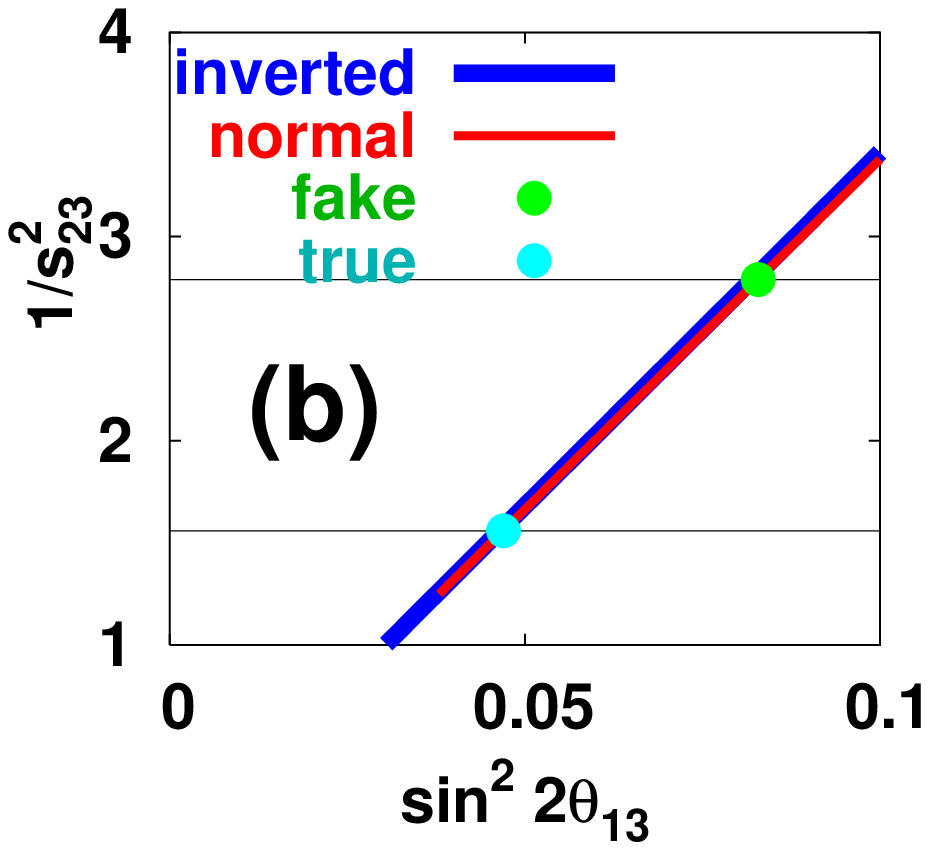}
\vglue 0.5cm
\caption{The $\theta_{23}$ ambiguity which could
arise after the JPARC measurements of
$P(\nu_\mu\rightarrow\nu_e)$,
$P(\bar{\nu}_\mu\rightarrow\bar{\nu}_e)$ and
$P(\nu_\mu\rightarrow\nu_\mu)$ at the oscillation
maximum.
(a) If $\sin^22\theta_{23}\simeq1.0$ then
the values of $\theta_{13}$ and $\theta_{23}$ are
close to each other for all the solutions, and the
ambiguity is not serious as far as $\theta_{13}$ and $\theta_{23}$ are
concerned.  (b) If $\sin^22\theta_{23}<1$ then
the $\theta_{23}$ ambiguity has to be resolved to determine
$\theta_{13}$ and $\theta_{23}$ to good precision.
}
\label{fig5}
\end{figure}

If $\sin^22\theta_{23}\simeq1$, then all the four solutions
are basically close to each other in the 
($\sin^22\theta_{13}$, $1/s^2_{23}$) plane, and the ambiguity
due to degeneracies are not serious as far as $\theta_{13}$ and
$\theta_{23}$ are concerned (See Fig.\,\ref{fig5}(a)).
On the other hand, if $\sin^22\theta_{23}$ deviates fairly from 1,
then the solutions are separated into two groups, those for
$\theta_{23}>\pi/4$ and those for $\theta_{23}<\pi/4$
in the ($\sin^22\theta_{13}$, $1/s^2_{23}$) plane, as is shown
in Fig.\,\ref{fig5}(b).  In this case resolution of the
$\theta_{23}\leftrightarrow\pi/2-\theta_{23}$ ambiguity is necessary
to determine $\theta_{13}$, $\theta_{23}$ and $\delta$.

\subsection{Fake effects on CP violation due to degeneracies}
\subsubsection{$\sin^22\theta_{23}\simeq1$}
If the JPARC experiment finds out
from the measurement of the disappearance probability
$P(\nu_\mu \to \nu_\mu)=P$ that $\sin^22\theta_{23}\simeq1.0$
with a good approximation, then we would not have to worry very much
about parameter degeneracy as far as $\theta_{13}$ and $\theta_{23}$
are concerned, since the values of $\theta_{13}$ and $\theta_{23}$
for all the different solutions are close to each other.

On the other hand, when it comes to the value of the CP phase
phase $\delta$, we have to be careful.  From Ref. \cite{Barger:2001yr}
the true value $\delta$ and the fake value $\delta'$ for the
CP phase satisfy the following:
\begin{eqnarray}
x^\prime\sin\delta^\prime &=&
x\sin\delta {f^2+\bar f^2-f\bar f\over f\bar f}
- {x^2\over\sin\Delta}{f^2+\bar f^2\over f\bar f}{f-\bar f\over 2yg} \,,
\label{xsind1}
\end{eqnarray}
where $x$, $y$ are given in Eq. (\ref{xy}), $f$, $\bar{f}$, $g$
are given in Eqs. (\ref{f}) and (\ref{g}), and $x'$ is defined by
\begin{eqnarray}
x^{\prime2} &=&
{x^2(f^2+\bar f^2-f\bar f)-2yg(f-\bar f)x\sin\delta\sin\Delta
\over f\bar f} \,.\nonumber
\end{eqnarray}
Eq.~(\ref{xsind1}) indicates that
even if $\sin\delta = 0$ we have
nonvanishing fake CP violating effect
\begin{eqnarray}
\sin\delta^\prime =
- x {f^2+\bar f^2\over f\bar f} {f - \bar f\over 2yg\sin\Delta}
\sqrt{f\bar f\over f^2+\bar f^2 - f\bar f} \,,
\label{xsind2}
\end{eqnarray}
if we fail to identify the correct sign of $\Delta m^2_{31}$.
In the case of the JPARC experiment, Eq. ~(\ref{xsind2}) implies
\begin{eqnarray}
\sin\delta^\prime \simeq-2.2\sin\theta_{13},
\nonumber
\end{eqnarray}
which is not negligible unless $\sin^22\theta_{13}\ll 10^{-2}$.
Therefore we have to know the sign of $\Delta m^2_{31}$
to determine the CP phase to good precision.

\subsubsection{$\sin^22\theta_{23}<1$}
As was explained in Sect. \ref{om}, if $\sin^22\theta_{23}$ deviates
fairly from 1, then we have to resolve the ambiguity
due to the $\theta_{23}$ degeneracy to determine the values
of $\theta_{13}$ and $\theta_{23}$.  As for the value of the CP phase
$\delta$, we can estimate how serious the effect of the $\theta_{23}$
ambiguity on the value of $\delta$ could be.
If the true value $\delta$ is zero, then the CP phase $\delta'$
for the fake solution can be estimated as \cite{Barger:2001yr}
\begin{eqnarray}
\sin2\theta_{13}^\prime \sin\delta^\prime &=&
\left|\frac{\Delta m^2_{21}}{\Delta m^2_{31}}\right|
\frac{g(f-\bar f)\sin2\theta_{12}}{f\bar f}
{\cot2\theta_{23} \over\sin\Delta} \,,\nonumber
\end{eqnarray}
where
\begin{eqnarray}
\sin^22\theta_{13}^{\prime} &=&
\sin^22\theta_{13} \tan^2\theta_{23}
+ \left(\frac{\Delta m^2_{21}}{\Delta m^2_{31}}\right)^2
{g^2 \sin^22\theta_{12} \over f\bar f}(1-\tan^2\theta_{23}),
\nonumber
\end{eqnarray}
and $f$, $\bar{f}$, $g$ are defined in Eqs. (\ref{f}) and (\ref{g}).
In the case of JPARC, we have
\begin{eqnarray}
\left|\sin\delta'\right|\sim\frac{1}{200}
\frac{|\cot2\theta_{23}|}{t_{23}}
\frac{1}{\sin2\theta_{13}}\lesssim
\frac{1}{500}\frac{1}{\sqrt{\sin^22\theta_{13}}},
\label{th23d}
\end{eqnarray}
where we have used the bound $0.90\le\sin^22\theta_{23}\le1.0$
from the atmospheric neutrino data in the second inequality,
so that we see that the ambiguity due to the $\theta_{23}$
does not cause a serious problem on determination of $\delta$
for $\sin^22\theta_{13}\gtrsim10^{-2}$.  It should be stressed,
however, that the effect on CP violation due to the sgn($\Delta m^2_{31}$)
ambiguity is also serious in this case.

\section{Resolution of ambiguities by the third
measurement after JPARC}
In this section, assuming that the JPARC experiment, which is expected
to be the
first superbeam experiment, measures
$P(\nu_\mu \to \nu_e)$ and
$P(\bar\nu_\mu \to \bar\nu_e)$
at the oscillation maximum $\Delta\equiv\Delta m^2_{31}L/4E=\pi/2$,
we will discuss how the third measurement after JPARC
can resolve the ambiguities by using the plot in the
($\sin^22\theta_{13}$, $1/s^2_{23}$) plane.
Resolution of the $\theta_{23}$ ambiguity has been discussed
using the disappearance measurement of
$P(\bar\nu_e \to \bar\nu_e)$ at reactors
\cite{Fogli:1996pv,Barenboim:2002nv,Minakata:2002jv,Huber:2003pm,Minakata:2003wq}
and the silver channel $P(\nu_e \to \nu_\tau)$ at
neutrino factories \cite{Donini:2002rm},
but it has not been discussed much using the
channel $\nu_\mu \to \nu_e$ \footnote{
There have been a lot of works \cite{degeneracies}
on how to resolve parameter degeneracies, but they discussed
mainly the intrinsic and sgn($\Delta m^2_{31}$) degeneracies,
and the present scenario, in which the third experiment
follows the JPARC results on
$P(\nu_\mu \to \nu_e)$ plus $P(\bar\nu_\mu \to \bar\nu_e)$
which are measured at the oscillation maximum, has not been considered.}.
Here we take the following reference values for the
oscillation parameters:
\begin{eqnarray}
&{\ }&\sin^22\theta_{12}=0.8,~~
\sin^22\theta_{13}=0.05,~~
\sin^22\theta_{23}=0.96,\nonumber\\
&{\ }&\Delta m^2_{21}=7\times10^{-5}\mbox{\rm eV}^2,~~
\Delta m^2_{31}=2.5\times10^{-3}\mbox{\rm eV}^2>0,~~
\delta=\pi/4.
\label{ref}
\end{eqnarray}

\subsection{$\nu_\mu\rightarrow\nu_e$}
\begin{figure}
\includegraphics[scale=1.0]{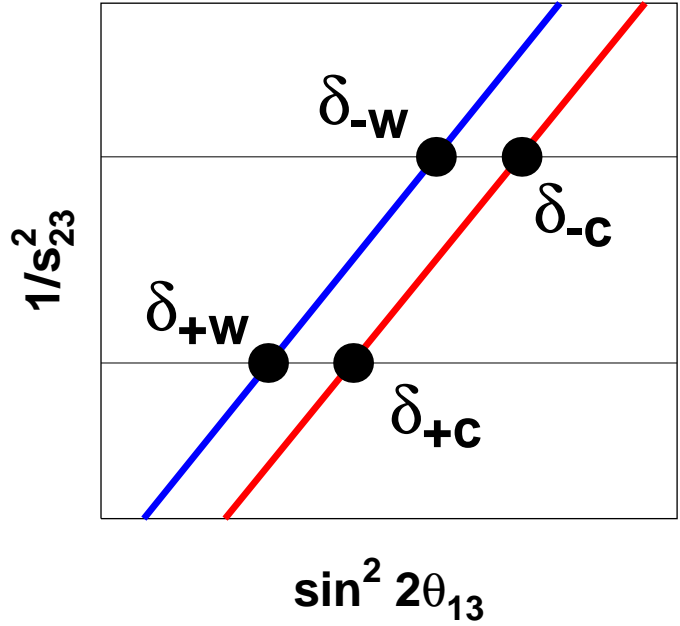}
\vglue 0.5cm
\caption{Four possible values for the CP phase $\delta$
at the oscillation maximum.  The red (blue) line stands for
the normal (inverted) hierarchy.  Since we are assuming
the normal hierarchy here, the red (blue) line corresponds
to the correct (wrong) assumption on the mass hierarchy.
$\pm$ sign stands for
the choice of $s^2_{23}=(1\pm\sqrt{1-\sin^22\theta_{23}})/2$
in the $\theta_{23}$ ambiguity, and c (w) stands for
the correct (wrong) assumption on the mass hierarchy.
}
\label{fig6}
\end{figure}

Let us discuss the case in which another long baseline
experiment measures $P(\nu_\mu \to \nu_e)$.
From the measurements of $P(\nu_\mu \to \nu_e)$ and
$P(\bar\nu_\mu \to \bar\nu_e)$ by JPARC at
the oscillation maximum we can deduce the 
value of $\delta$, up to the eightfold ambiguity
($\delta\leftrightarrow\pi-\delta$,
$\theta_{23}\leftrightarrow\pi/2-\theta_{23}$,
$\Delta m^2_{31}\leftrightarrow-\Delta m^2_{31}$).\footnote{
I thank Hiroaki Sugiyama for pointing this out to me.}
As is depicted in Fig.\,\ref{fig6}, depending on whether
$s^2_{23}-1/2$ is positive or negative, we assign the
subscript $\pm$, and depending on whether our ansatz
for sgn($\Delta m^2_{31}$) is correct or wrong, we assign
the subscript c or w.  Thus the eight possible
values of $\delta$ are given by
\begin{eqnarray}
\delta_{+\text{w}}, \delta_{+\text{c}},
\delta_{-\text{w}}, \delta_{-\text{c}},
\pi-\delta_{+\text{w}}, \pi-\delta_{+\text{c}},
\pi-\delta_{-\text{w}}, \pi-\delta_{-\text{c}}.
\label{8delta}
\end{eqnarray}
Now suppose that the third measurement gives the value $P$
for the oscillation probability $P(\nu_\mu \to \nu_e)$.
Then there are in general eight lines
in the ($X\equiv\sin^22\theta_{13}$, $Y\equiv 1/s^2_{23}$) plane
given by
\begin{eqnarray}
f^2X&=&\left[P-C+2C\cos^2(\delta+\Delta)\right](Y-1)+P\nonumber\\
&{\ }&-2\cos(\delta+\Delta)\sqrt{C(Y-1)}
\sqrt{\left[P-C\sin^2(\delta+\Delta)\right](Y-1)+P}
\label{p3n}
\end{eqnarray}
for the normal hierarchy, and
\begin{eqnarray}
\bar{f}^2X&=&\left[P-C+2C\cos^2(\delta-\Delta)\right](Y-1)+P\nonumber\\
&{\ }&-2\cos(\delta-\Delta)\sqrt{C(Y-1)}
\sqrt{\left[P-C\sin^2(\delta-\Delta)\right](Y-1)+P}
\label{p3i}
\end{eqnarray}
for the inverted hierarchy, where $C$ is defined in Eq. (\ref{c}),
$\Delta\equiv|\Delta m^2_{31}|L/4E$ is defined for the third measurement,
and $\delta$ takes one of the eight values given in Eq. (\ref{8delta}).
The derivation of (\ref{p3n}) and (\ref{p3i}) is given in
Appendix \ref{appendix1}.
It turns out that the solutions (\ref{p3n}) and (\ref{p3i}) are
hyperbola if $\cos^2(\delta\pm\Delta)>(C-P)/P$,
where $+$ and $-$ refer to the normal and inverted hierarchy,
and ellipses if $\cos^2(\delta\pm\Delta)<(C-P)/P$.
In practice, however, the difference between hyperbola and
ellipses is not so important for the present discussions,
because we are only interested in the behaviors of these curves
in the region $1.52<Y\equiv 1/s^2_{23}<2.92$ which comes
from the 90\%CL allowed region of the Superkamiokande atmospheric
neutrino data for $\sin^22\theta_{23}$.

Here let us look at three typical cases:
$L$=295km, $L$=730km, $L$=3000km, each of which corresponds
to JPARC, off-axis NuMI \cite{Ayres:2002nm},
and a neutrino factory \cite{Geer:1997iz} \footnote{
For $L$=3000km the density of the matter may not be treated as constant,
and the probability formulae (\ref{degene3n}) and (\ref{degene3i})
may no longer be valid.  It turns out, however, that the
approximation of the formulae becomes good if we replace
$AL$ by $AL\rightarrow\int^L_0A(x)dx$ everywhere in the formula.
In the following discussions, the replacement
$AL\rightarrow\int^L_0A(x)dx$ is always understood in the case of the
baseline $L$=3000km.  It should be mentioned that
the neutrino energy spectrum at neutrino factories is
continuous and it is assumed here that we take one particular
energy bin whose energy range can be made relatively small.
It should be also noted that neutrino factories actually
measure the probabilities $P(\nu_e \to \nu_\mu)$ or
$P(\bar{\nu}_e \to \bar{\nu}_\mu)$, instead of
$P(\nu_\mu \to \nu_e)$ or
$P(\bar{\nu}_\mu \to \bar{\nu}_e)$.  Here we discuss for simplicity
the trajectory of $P(\nu_\mu \to \nu_e)$ whose feature
is the same as that of $P(\nu_e \to \nu_\mu)$.}

\begin{figure}
\vglue -1.0cm
\includegraphics[width=6.5cm]{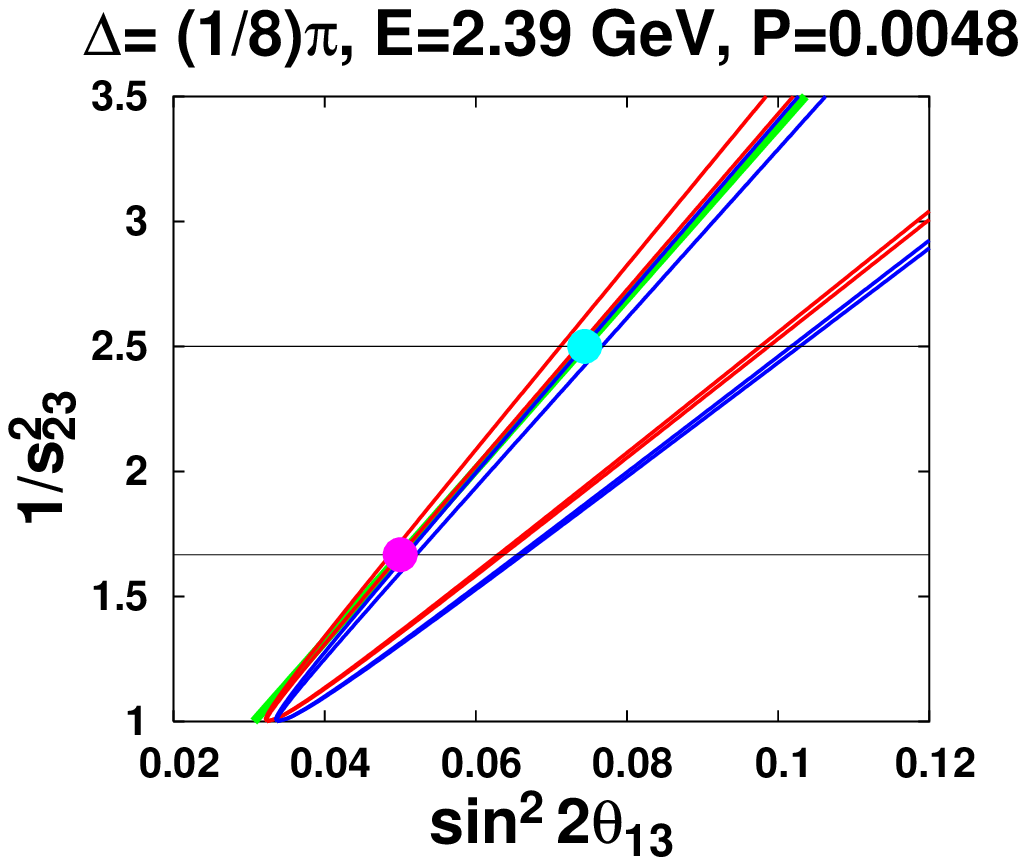}
\includegraphics[width=6.5cm]{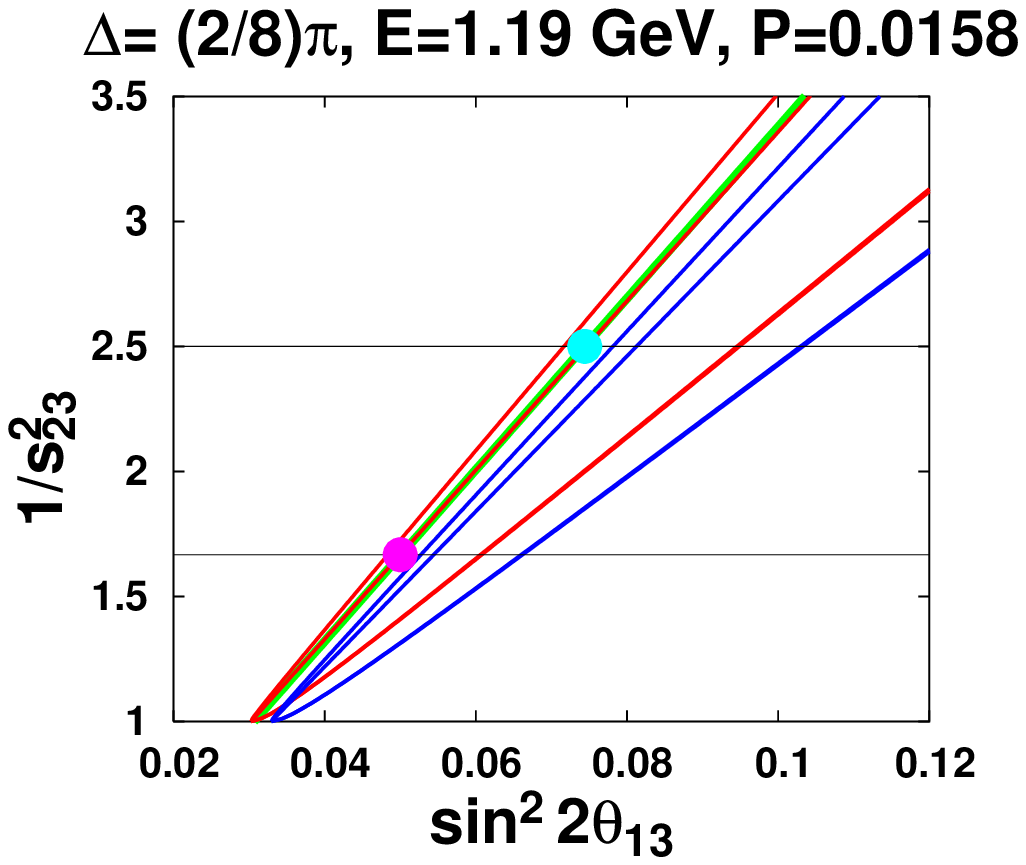}
\includegraphics[width=6.5cm]{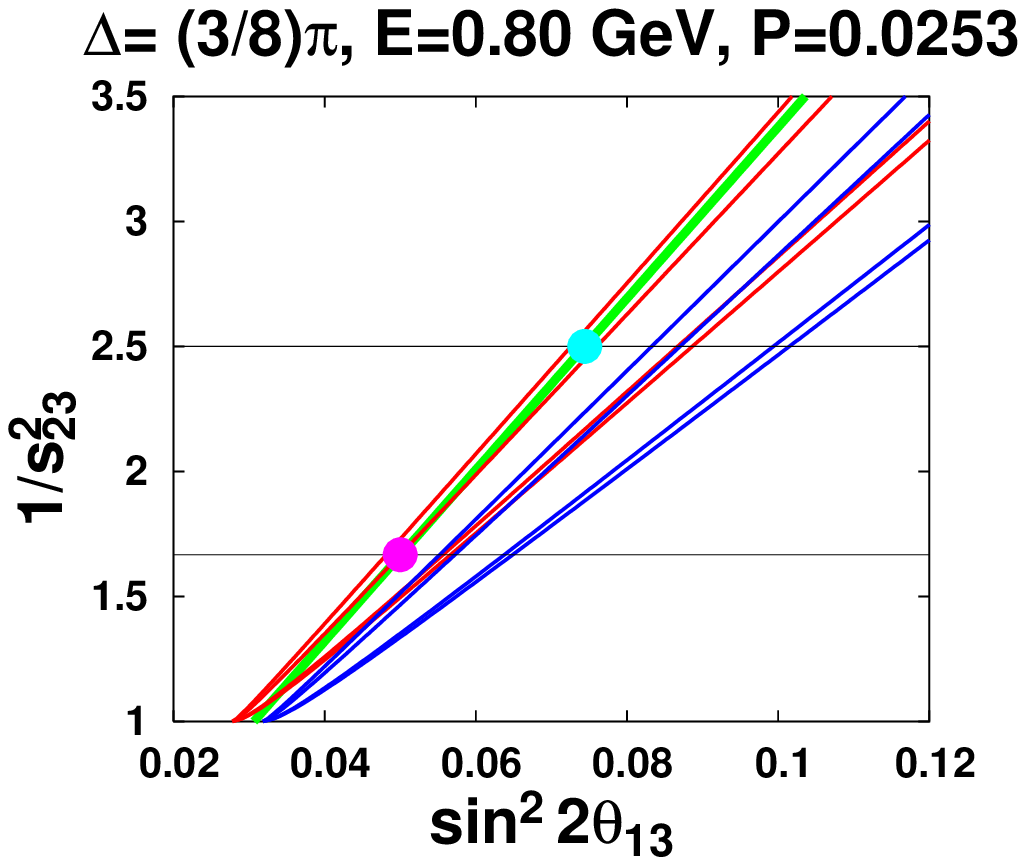}
\includegraphics[width=6.5cm]{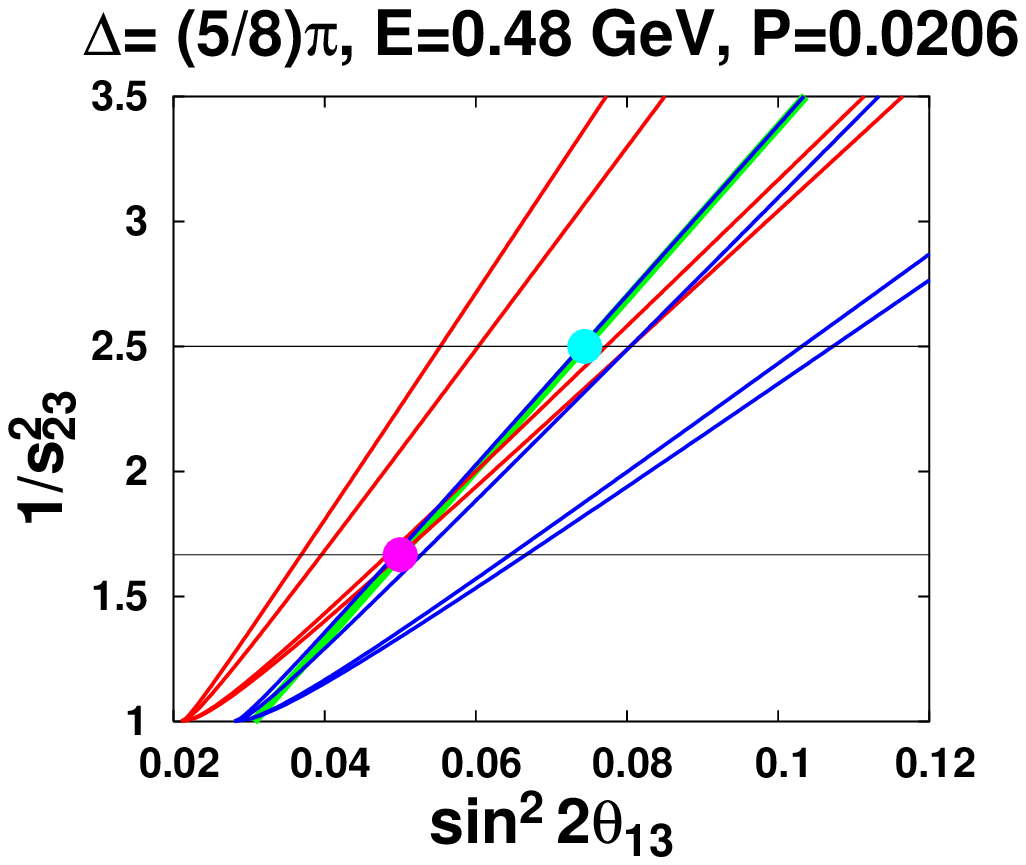}
\includegraphics[width=6.5cm]{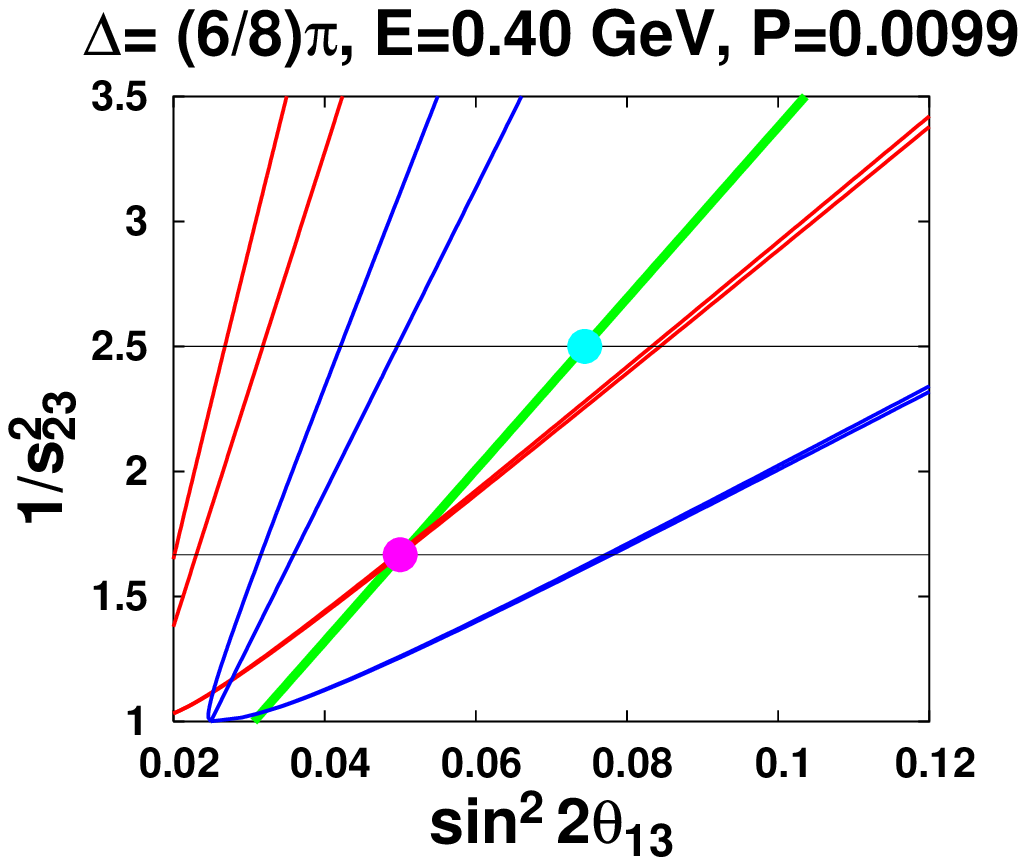}
\includegraphics[width=6.5cm]{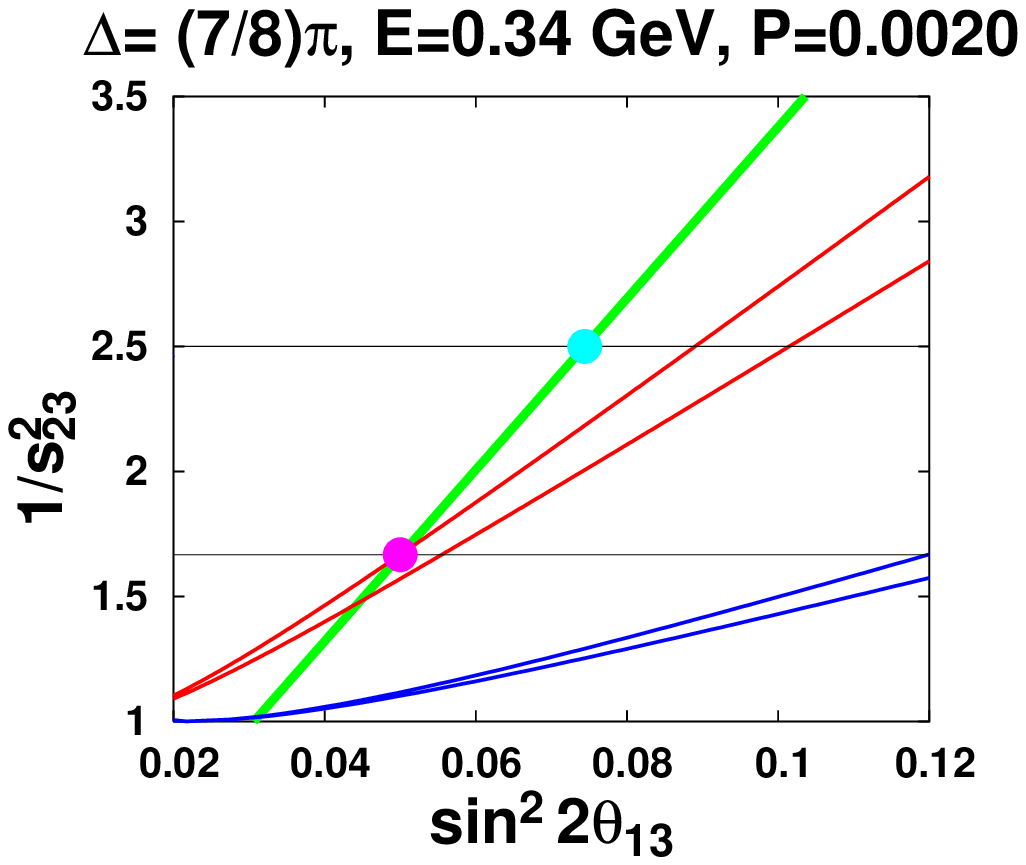}
\includegraphics[width=6.5cm]{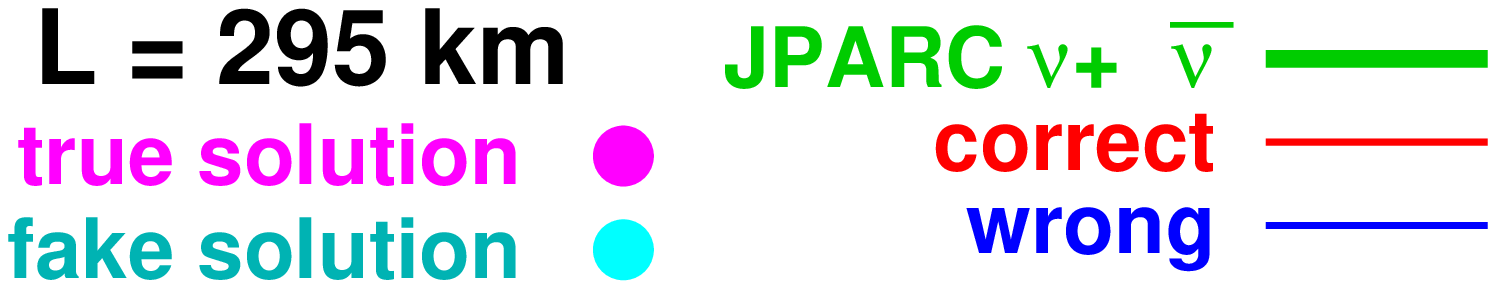}
\vglue -3cm 
\caption{\small
The trajectories of $P(\nu_\mu\rightarrow\nu_e)=$ const.
of the third experiment at $L$=295km with
$\Delta\equiv|\Delta m^2_{31}|L/4E=(j/8)\pi~(0\le j\le 7,j\ne4)$
after JPARC.  The true values are those in Eq. (\ref{ref}).
The green line is the JPARC result
obtained by $P(\nu_\mu\rightarrow\nu_e)$ and
$P(\bar{\nu}_\mu\rightarrow\bar{\nu}_e)$ at the oscillation
maximum.  The red (blue) lines are the trajectories
of $P(\nu_\mu\rightarrow\nu_e)$ given by the third experiment
assuming the normal (inverted) hierarchy, where $\delta$ takes
four values for each mass hierarchy.
}
\label{fig7}
\end{figure}

\begin{figure}
\vglue -1.0cm
\includegraphics[width=6.5cm]{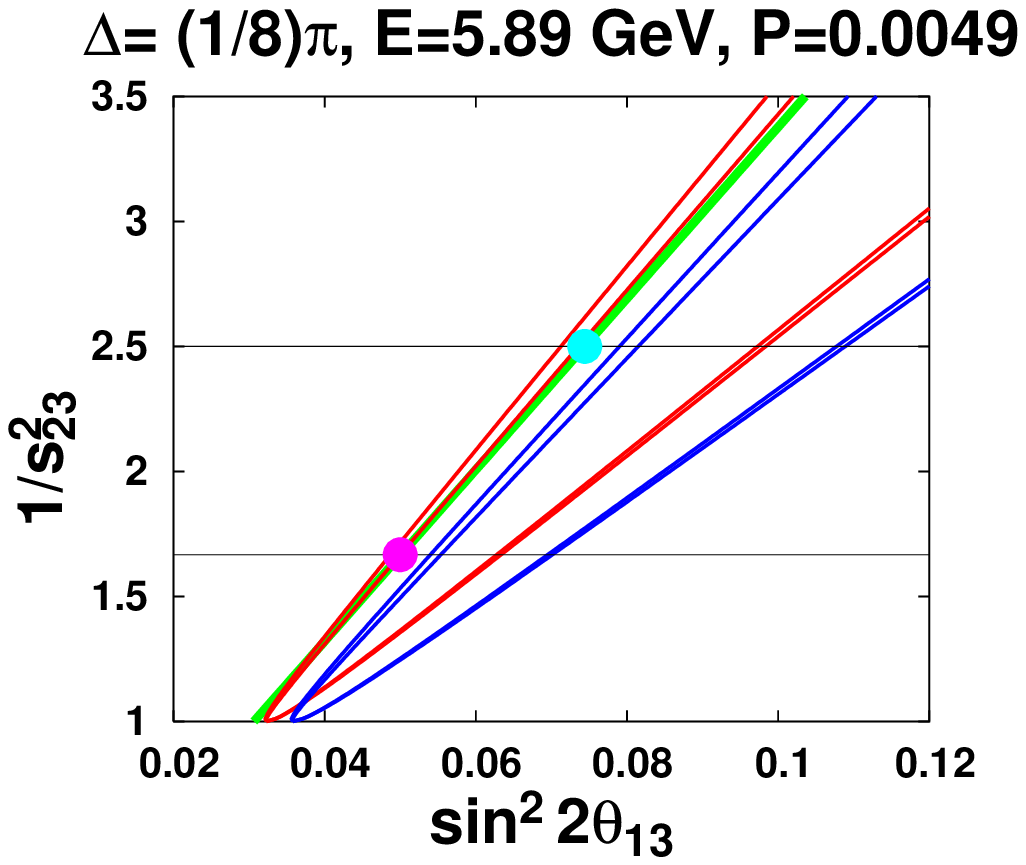}
\includegraphics[width=6.5cm]{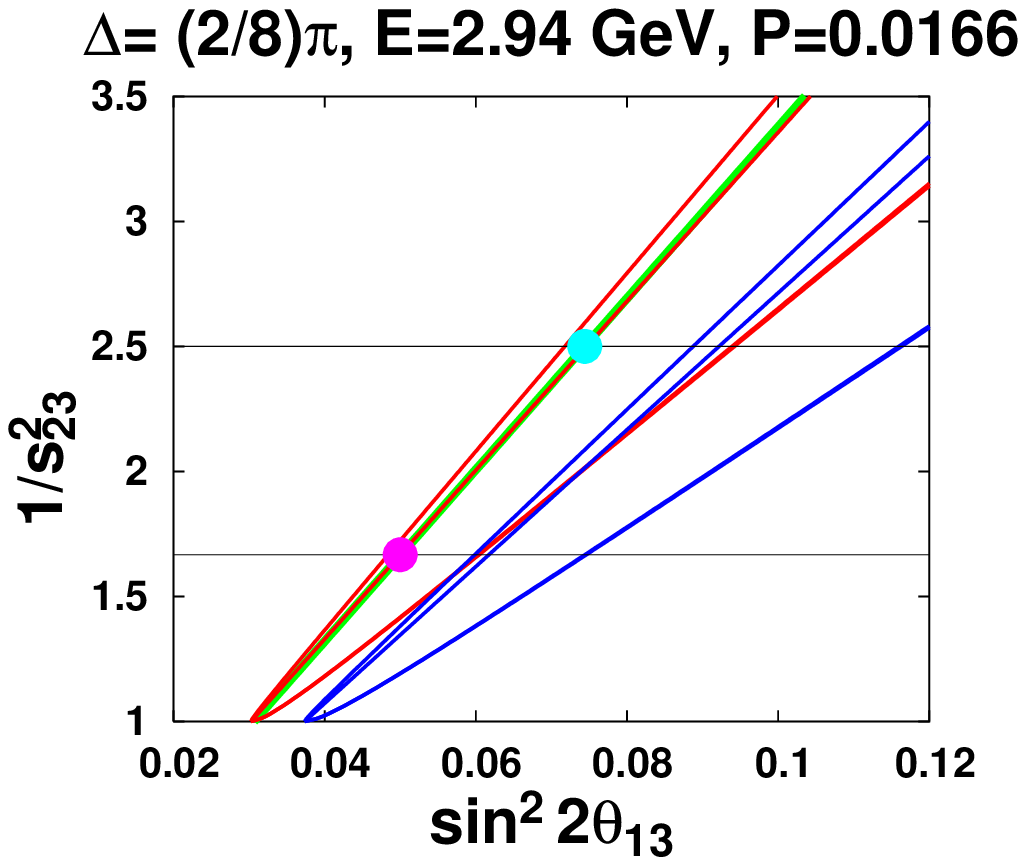}
\includegraphics[width=6.5cm]{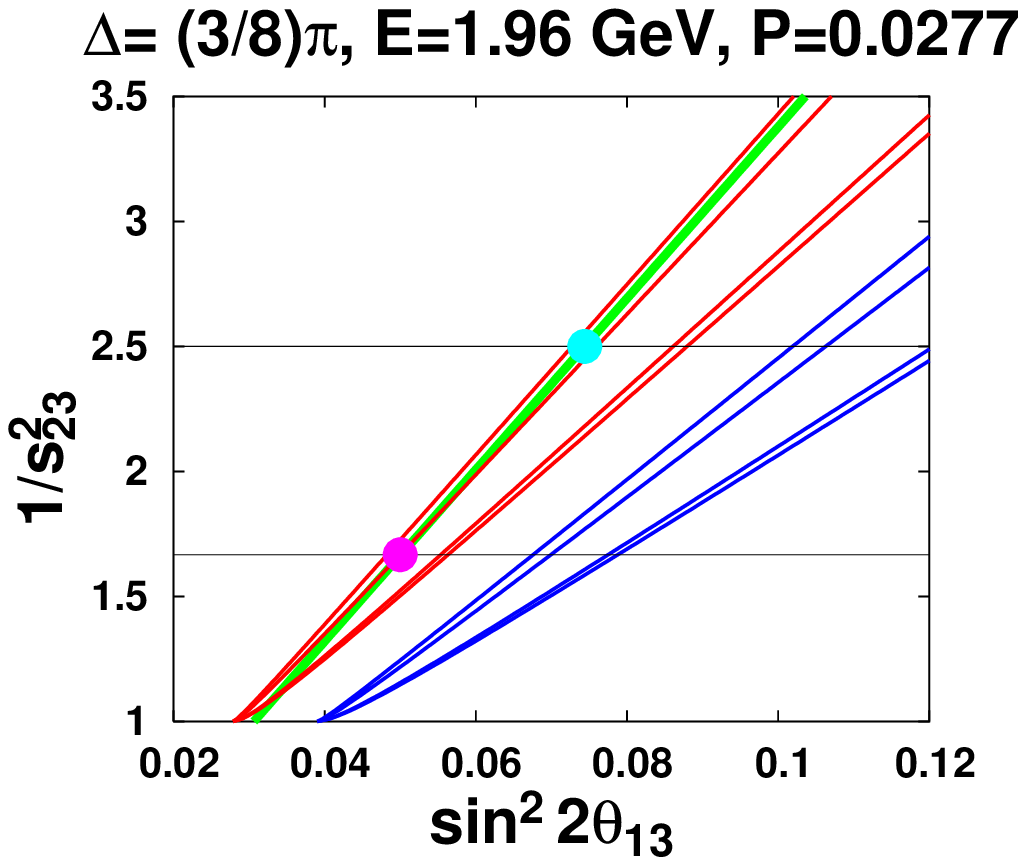}
\includegraphics[width=6.5cm]{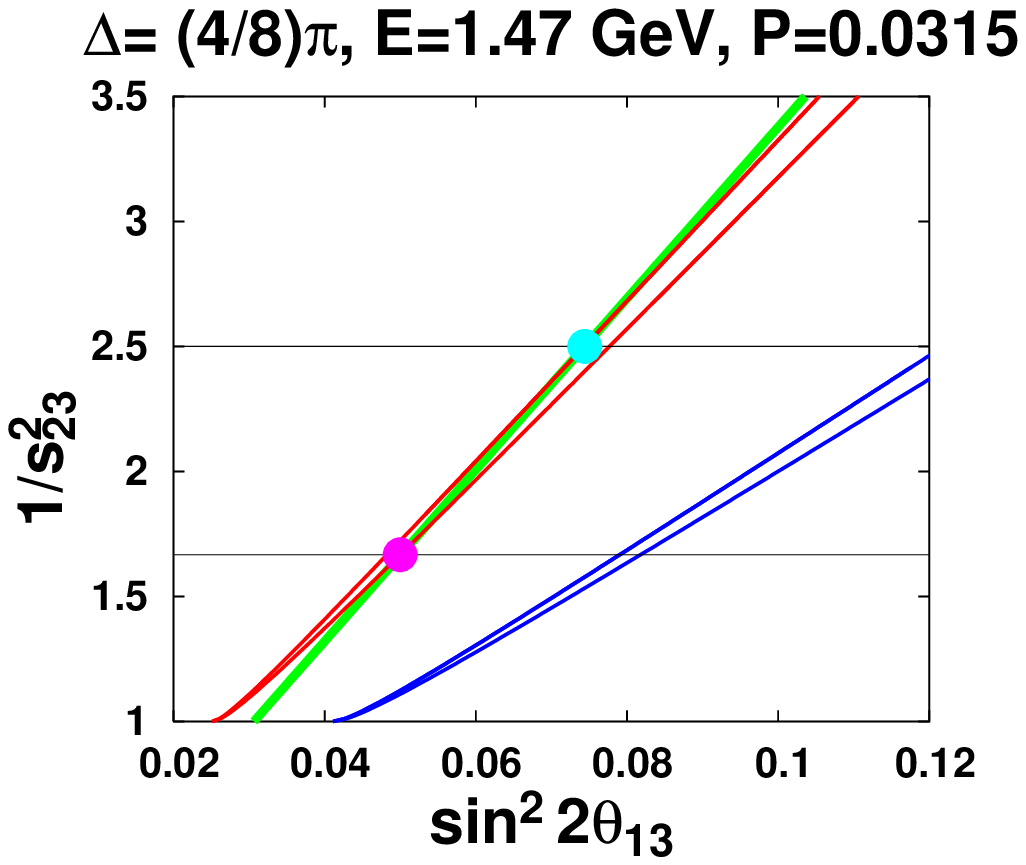}
\includegraphics[width=6.5cm]{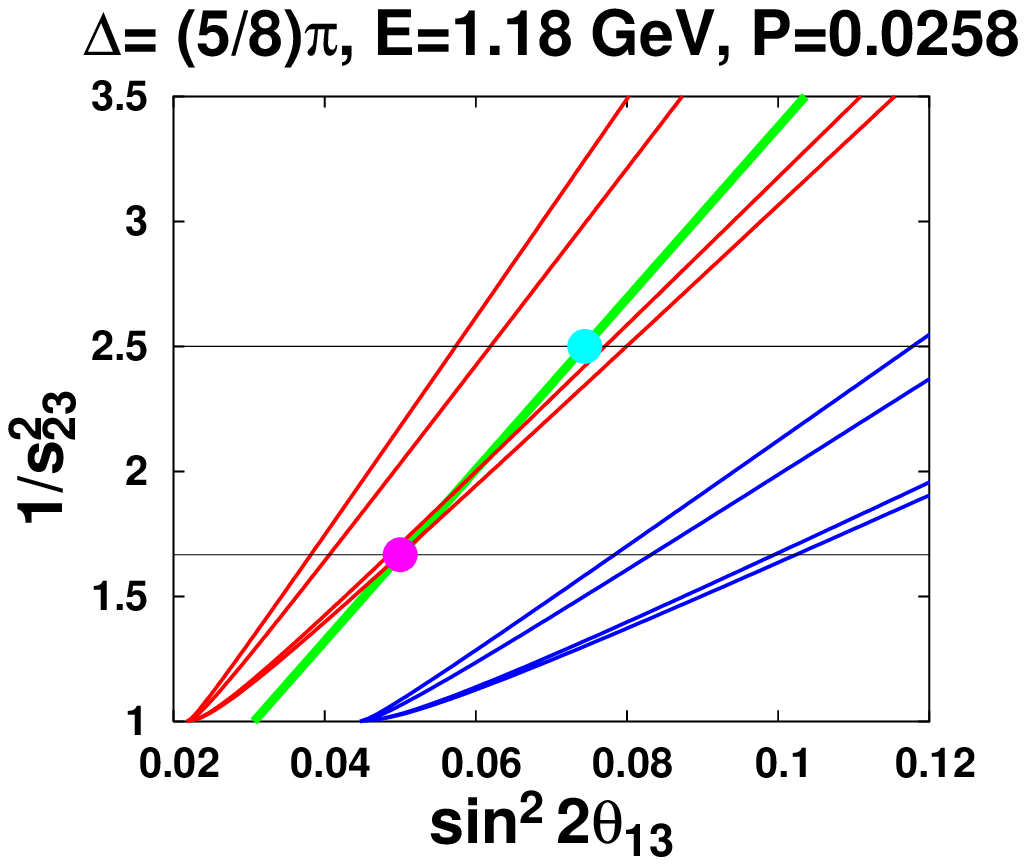}
\includegraphics[width=6.5cm]{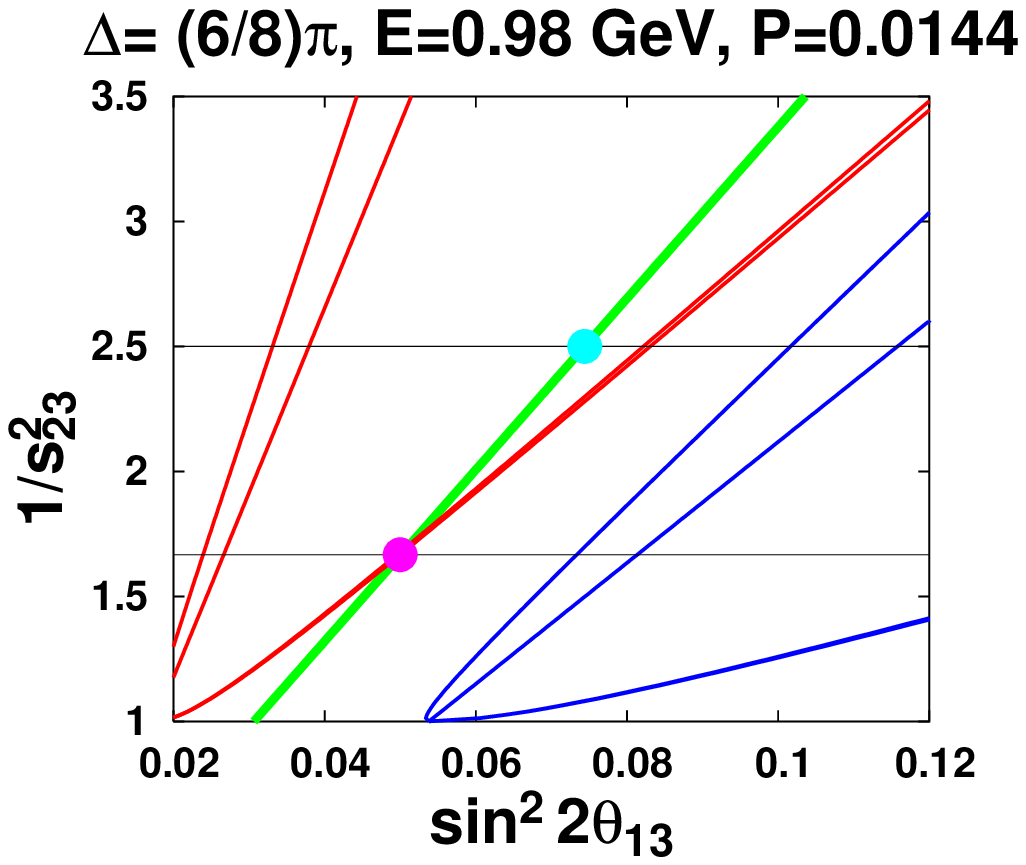}
\includegraphics[width=6.5cm]{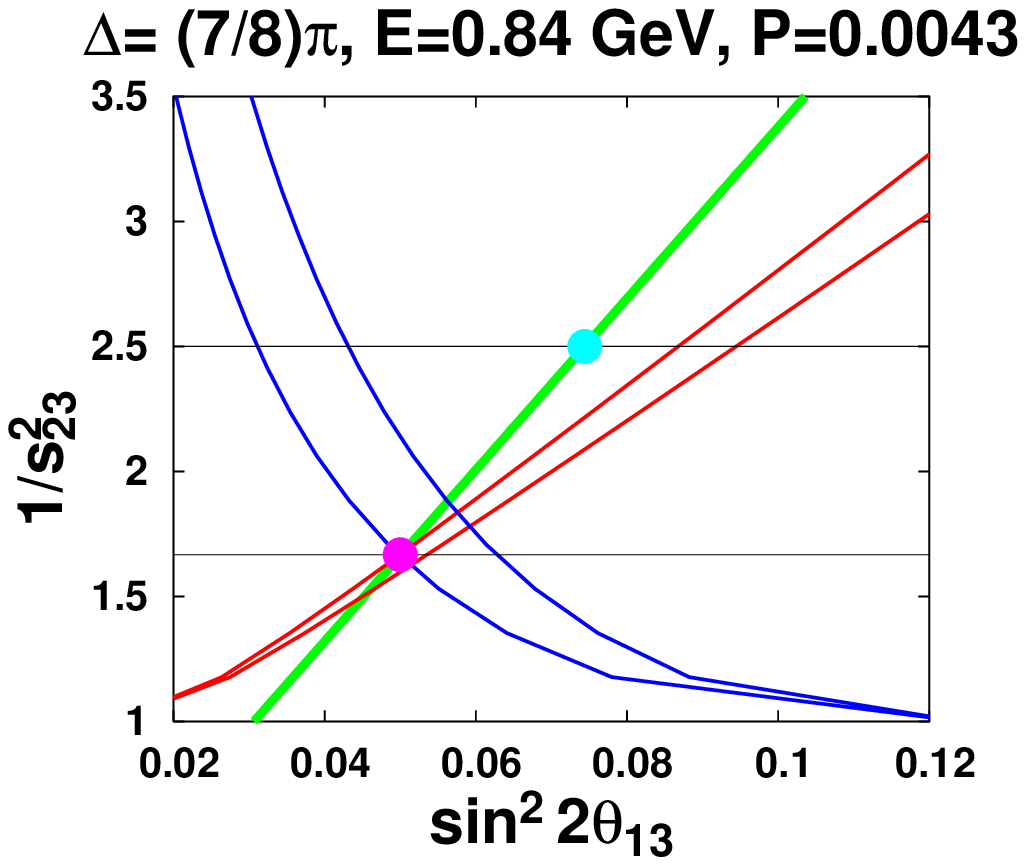}
\includegraphics[width=6.5cm]{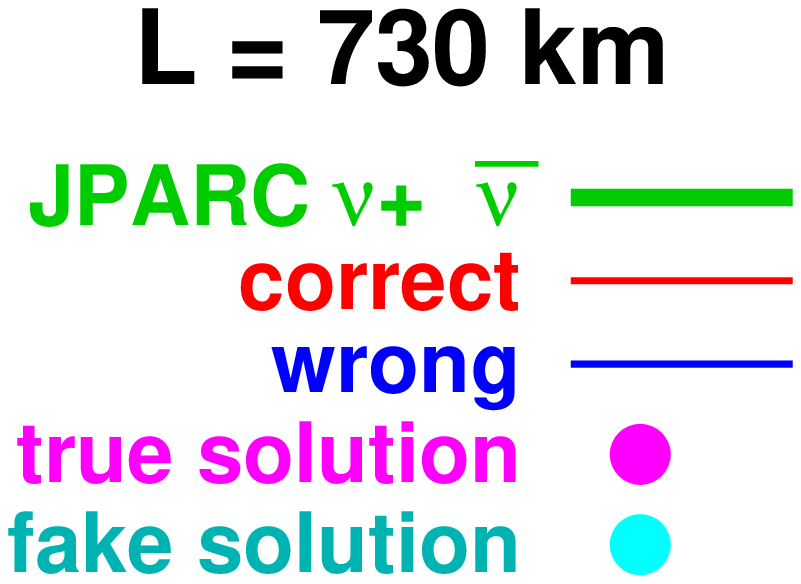}
\caption{\small
The trajectories of $P(\nu_\mu\rightarrow\nu_e)=$ const.
of the third experiment at $L$=730km with
$\Delta\equiv|\Delta m^2_{31}|L/4E=(j/8)\pi~(0\le j\le 7)$ after JPARC.
The true values are those in Eq. (\ref{ref}).}
\label{fig8}
\vglue -0.5cm 
\end{figure}

\begin{figure}
\vglue -1.5cm
\includegraphics[width=6.5cm]{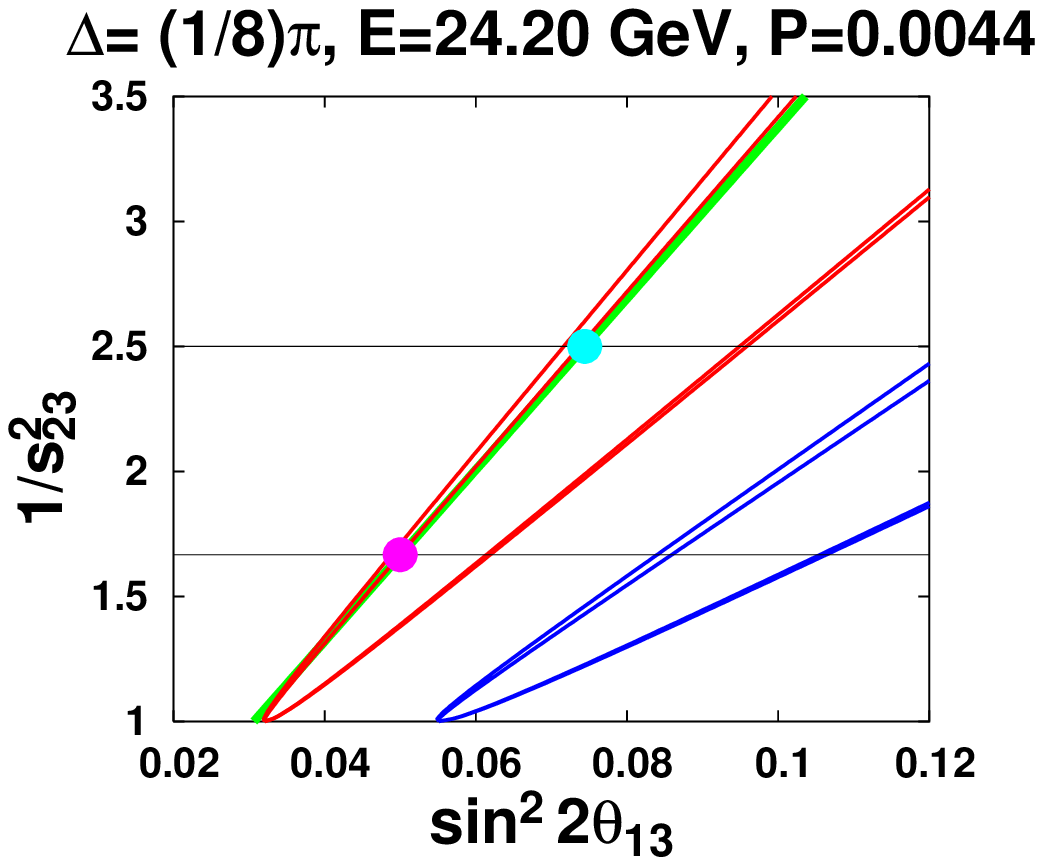}
\includegraphics[width=6.5cm]{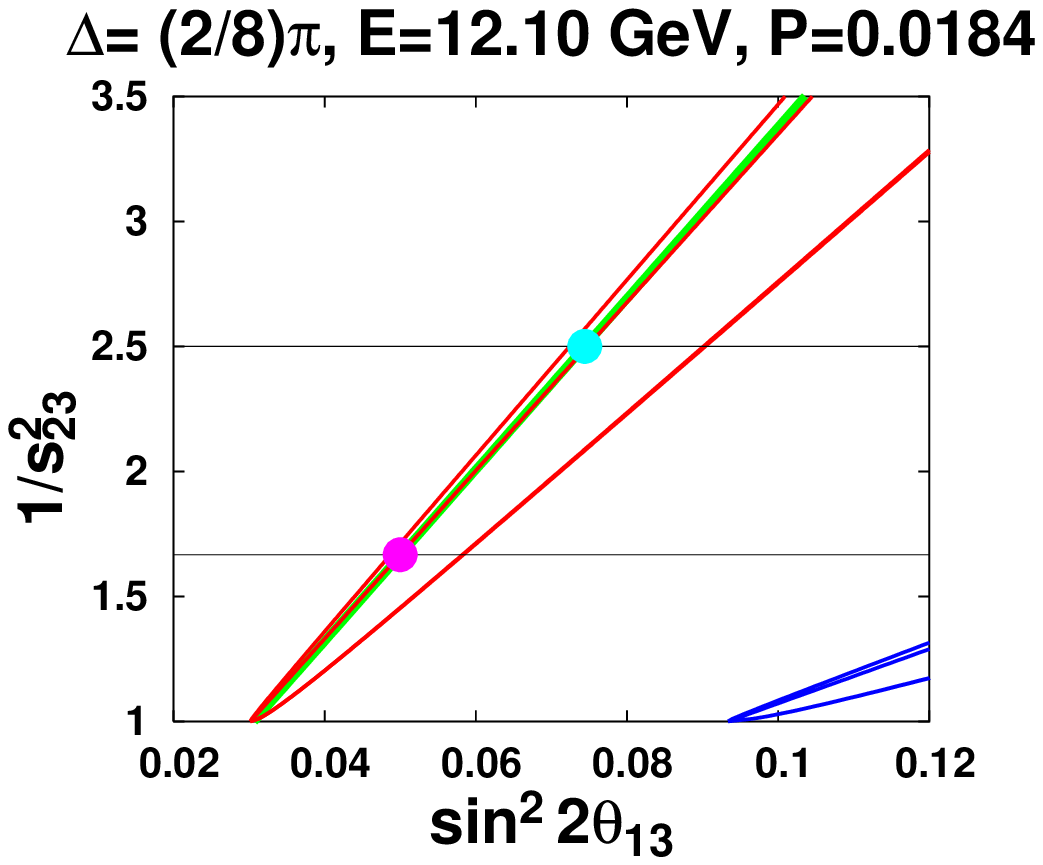}
\includegraphics[width=6.5cm]{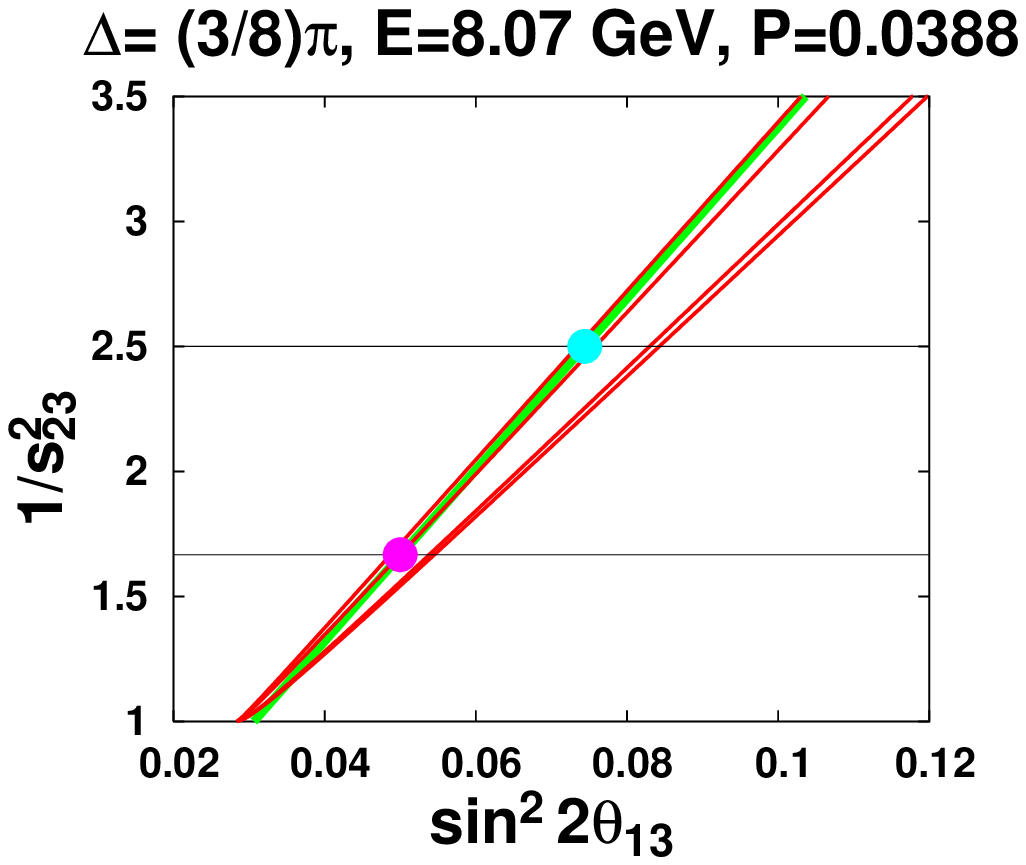}
\includegraphics[width=6.5cm]{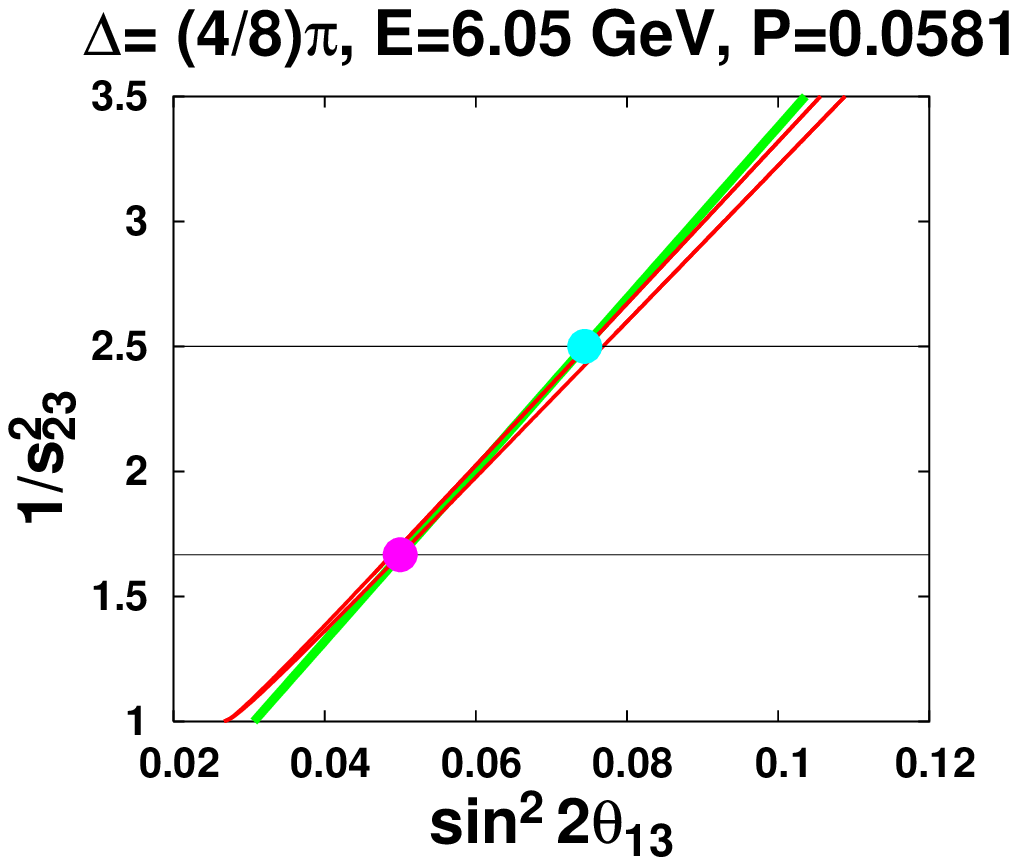}
\includegraphics[width=6.5cm]{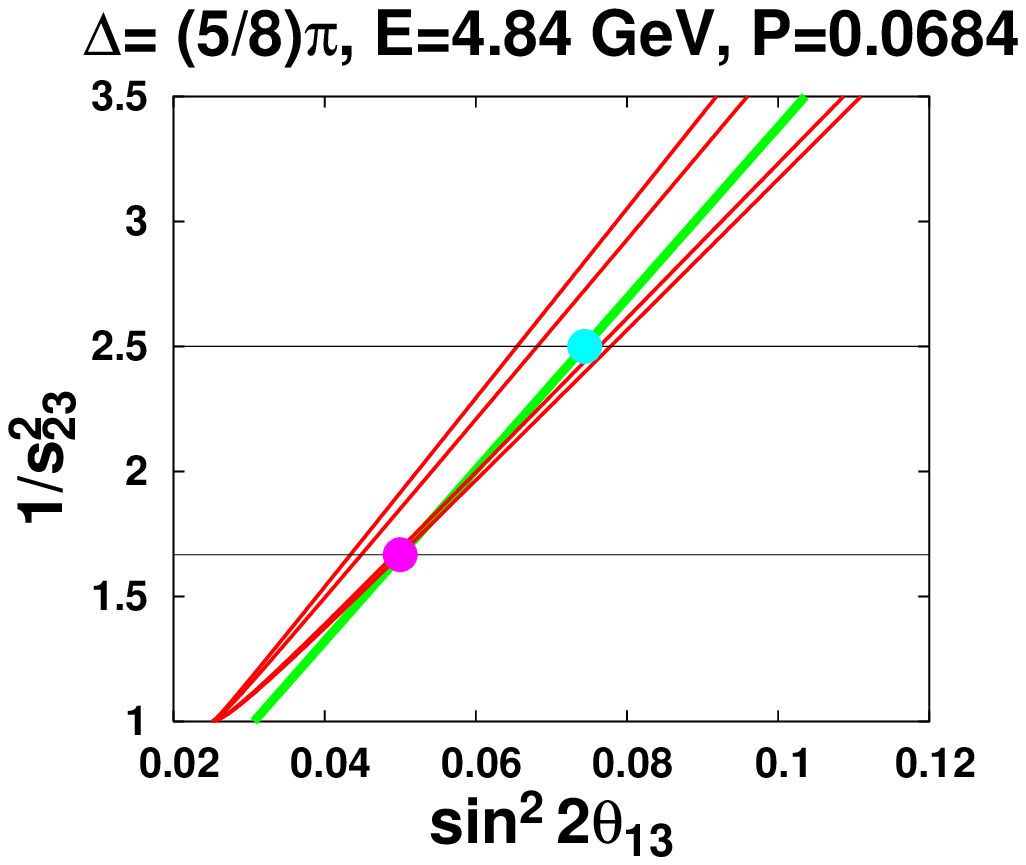}
\includegraphics[width=6.5cm]{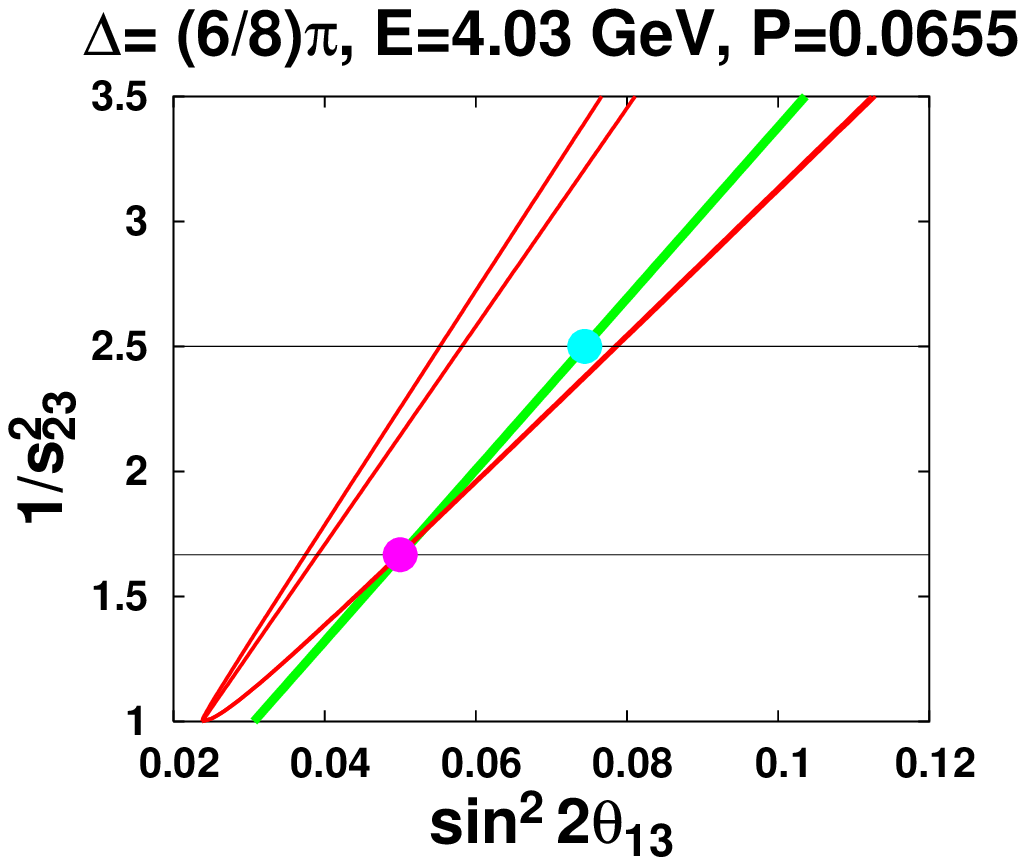}
\includegraphics[width=6.5cm]{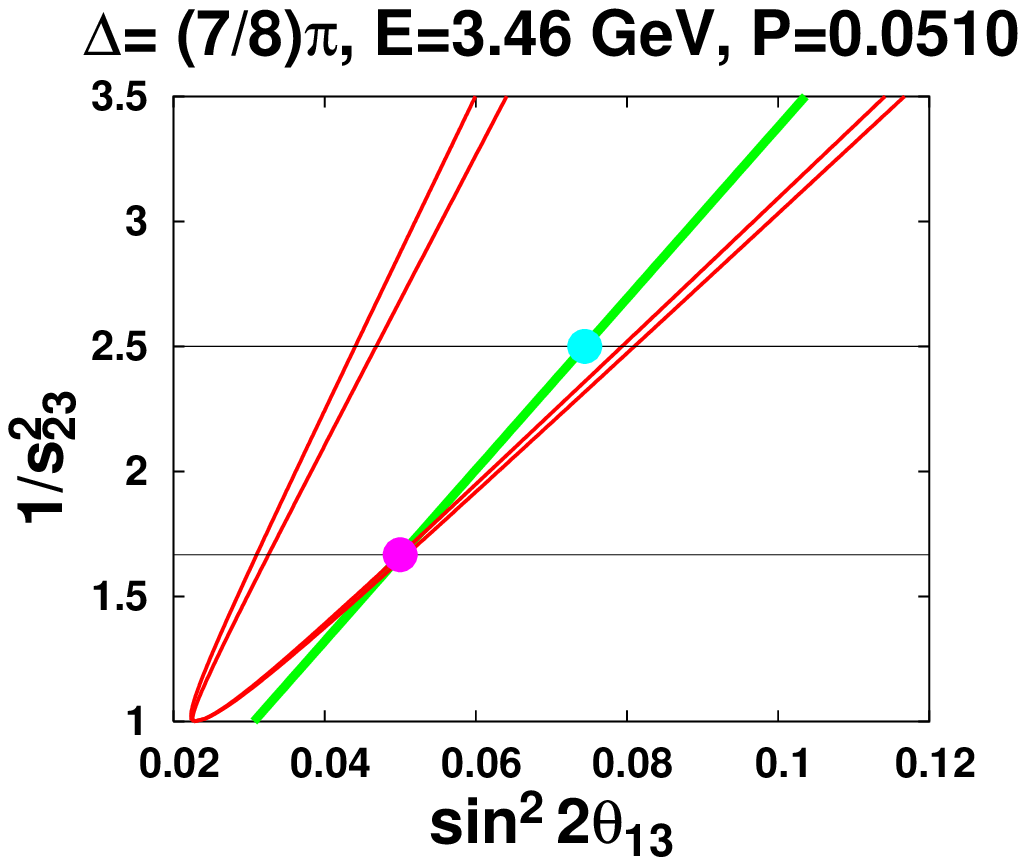}
\includegraphics[width=6.5cm]{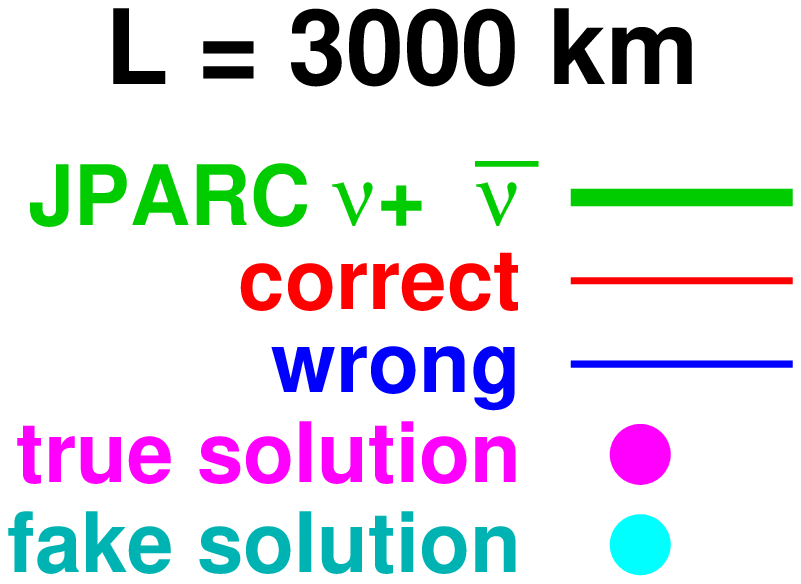}
\caption{\small
The trajectories of $P(\nu_\mu\rightarrow\nu_e)=$ const.
of the third experiment at $L$=3000km with
$\Delta\equiv|\Delta m^2_{31}|L/4E=(j/8)\pi~(0\le j\le 7)$ after JPARC.
The true values are those in Eq. (\ref{ref}).
For $\Delta\ge(3/8)\pi$, the blue curves (with the wrong assumption
for the mass hierarchy) are not in the figure because they are far to the right.}
\label{fig9}
\vglue -0.5cm 
\end{figure}

Figs.\,\ref{fig7},\ref{fig8},\ref{fig9} show
the trajectories of $P(\nu_\mu \to \nu_e)$
obtained in the third measurement together with
the constraint of $P(\nu_\mu \to \nu_e)$,
$P(\bar\nu_\mu \to \bar\nu_e)$ and $P(\nu_\mu \to \nu_\mu)$
by JPARC, for $L$=295km, $L$=730km, $L$=3000km, respectively,
where $\Delta\equiv|\Delta m^2_{31}|L/4E$ takes the values
$\Delta=j\pi/8$ ($j=1,\cdots,7$).
The purple (light blue) blob stands for the
true (fake) solution given by the JPARC results on
$P(\nu_\mu \to \nu_e)$,
$P(\bar\nu_\mu \to \bar\nu_e)$ and $P(\nu_\mu \to \nu_\mu)$.
For the correct (wrong) guess on the mass hierarchy,
there are in general four red (blue) curves because
the CP phase $\delta$, which is deduced from the JPARC results on
$P(\nu_\mu \to \nu_e)$,
$P(\bar\nu_\mu \to \bar\nu_e)$ and $P(\nu_\mu \to \nu_\mu)$,
is fourfold:
($\delta_{+\text{c}}$, $\delta_{-\text{c}}$,
$\pi-\delta_{+\text{c}}$, $\pi-\delta_{-\text{c}}$)
for the correct assumption on the hierarchy and
($\delta_{+\text{w}}$, $\delta_{-\text{w}}$, 
$\pi-\delta_{+\text{w}}$, $\pi-\delta_{-\text{w}}$)
for the wrong assumption.
In most cases the four (red or blue) curves are separated into two pairs of
curves.  As we will see later, the large split is due to
the $\delta\leftrightarrow\pi-\delta$ ambiguity, while
the small split is due to the $\theta_{23}\leftrightarrow\pi/2-\theta_{23}$
ambiguity.  The reason that the latter splitting is small is because
the difference of the values in the CP phases is small, as is seen from
Eq. (\ref{th23d}).
In some of the figures in Figs.\,\ref{fig7},\ref{fig8},\ref{fig9}
the number of the red or blue curves is less than four
because not all values of $\delta$ give consistent solutions
for a set of the oscillation parameters.

Let us discuss each ambiguity one by one.

\subsubsection{$\delta\leftrightarrow\pi-\delta$ ambiguity}
As was mentioned above, the large splitting
of four (red or blue) lines into two pair of lines is due to
the $\delta\leftrightarrow\pi-\delta$ ambiguity.
From Eqs. (\ref{p3n}) and (\ref{p3i}) we see that the only
difference of the solutions with $\delta$ and with $\pi-\delta$
appears in $\cos(\delta\pm\Delta)$ or $\sin(\delta\pm\Delta)$.
If $\Delta=\pi/2$ (i.e., the oscillation maximum), we have
$\cos(\delta+\Delta)=-\sin\delta$,
$\cos(\pi-\delta+\Delta)=-\sin\delta$, so that the values
of $X$ with $\delta$ and with $\pi-\delta$ are the same, i.e.,
at oscillation maximum there is exact $\delta\leftrightarrow\pi-\delta$
degeneracy.  On the other hand, if $\Delta\ne\pi/2$, we have
$\cos(\delta+\Delta)\ne\cos(\pi-\delta+\Delta)$, and the
values of $X$ with $\delta$ and with $\pi-\delta$ are different.
Thus, to resolve the $\delta\leftrightarrow\pi-\delta$ ambiguity
it is advantageous to perform an experiment at $\Delta$ which
is farther away from $\pi/2$.  Deviation of $\Delta$ from $\pi/2$
implies either high energy or low energy.  In general
the number of events increases for high energy because
both the cross section and the neutrino flux increase, so
the high energy option is preferred to
resolve the $\delta\leftrightarrow\pi-\delta$ ambiguity \footnote{
Resolution of $\delta\leftrightarrow\pi-\delta$ ambiguity at
neutrino factories was discussed in \cite{Pinney:2001xw}}.

\subsubsection{$\Delta m^2_{31}\leftrightarrow-\Delta m^2_{31}$ ambiguity}
As one can easily imagine, the sgn($\Delta m^2_{31}$)
ambiguity is resolved better with longer baselines, since
the dimensionless quantity
$AL\equiv\sqrt{2}G_FN_eL\sim(L/1900$km)($\rho/2.7$g$\cdot$cm$^{-3})$
becomes of order one for $L\gtrsim$1000km.
On the other hand, from Figs.\,\ref{fig8} and \ref{fig9},
we observe that the split of the curves with the
different mass hierarchies
(the red vs blue curves) is larger for lower energy.
Naively this appears to be counterintuitive, because
at low energy the matter effect is expected to be
less important ($|\Delta m^2_{31}|L/4E\gg AL$).
However, this is not the case because we are dealing with
the value of $\sin^22\theta_{13}$ which is obtained for a given value of
$P(\nu_\mu \to \nu_e)$.  To see this, let us consider for simplicity
the the value of $X\equiv\sin^22\theta_{13}$ at $Y\equiv1/s^2_{23}=1$,
i.e., the X-intercept of the quadratic curves at $Y=1$.
$(\sin^22\theta_{13})_n$ ($(\sin^22\theta_{13})_i$) at $Y=1$ for the
normal (inverted) hierarchy is given by $x^2$
by putting $y=0$ in Eq. (\ref{degene3n}) (Eq. (\ref{degene3i})):
\begin{eqnarray}
(\sin^22\theta_{13})_n&=&\frac{P}{f^2}\nonumber\\
&{\ }&\qquad\qquad\qquad\mbox{\rm for}~s^2_{23}=1\nonumber\\
(\sin^22\theta_{13})_i&=&\frac{P}{\bar{f}^2}.
\nonumber
\end{eqnarray}
The ratio of these two quantities is given for small $AL$ by
\begin{eqnarray}
\frac{(\sin^22\theta_{13})_n}{(\sin^22\theta_{13})_i}&=&
\frac{f^2}{\bar{f}^2}=\frac{\sin^2\left(\Delta-AL/2\right)}
{\sin^2\left(\Delta+AL/2\right)}
\left(\frac{1+AL/2\Delta}{1-AL/2\Delta}\right)^2\nonumber\\
&\simeq&1+2AL\left(\frac{1}{\Delta}-\frac{1}{\tan\Delta}\right),
\nonumber
\end{eqnarray}
so that the larger $\Delta$ is (the smaller the neutrino energy is),
the larger this ratio becomes, as long as $\Delta$ does not exceed $\pi$.
This phenomenon suggests that it is potentially possible to
enhance the matter effect by performing an experiment at low energy
($\Delta>\pi/2$) even with $L$=730km, and it may enable us to
determine the sign of $\Delta m^2_{31}$ at the off-axis NuMI experiment.
While the neutrino flux decreases for low energy at the off-axis NuMI
experiment, the cross section at $E\sim$1GeV is not particularly small
compared to higher energy, so the low energy possibility at
the off-axis NuMI experiment deserves serious study.

\subsubsection{$\theta_{23}\leftrightarrow\pi/2-\theta_{23}$ ambiguity}
Figs.\,\ref{fig7},\ref{fig8},\ref{fig9}, which are plotted for
$\delta=\pi/4$,
suggest that there is a tendency in which the slope of the red curve
which goes through the true point (the purple blob) is almost the
same for high energy as that of the straight green line obtained by JPARC,
while
for the low energy the slope of the red curve is smaller than that of
the JPARC green line.  Here we will discuss the $X$-intercept at
$Y=1$ instead of calculating the slope itself, because it is easier
to consider the $X$-intercept and because the difference in the
$X$-intercepts inevitably implies the different slopes for the two
lines, as almost all the curves are approximately straight lines.
In the case of JPARC, the matter effect is small ($AL\simeq0.08$)
so that we can put $f\simeq\bar{f}\simeq1$.
From Eq. (\ref{omn}) we have the $X$-intercept at $Y=1$
\begin{eqnarray}
X_\text{JPARC}=\frac{P/f+\bar{P}/\bar{f}}{f+\bar{f}}
\simeq\frac{P+\bar{P}}{2}\simeq x^2,
\label{xjparc}
\end{eqnarray}
where the term $g^2y^2$ has been ignored for simplicity.
On the other hand, for the third measurement, from Eq. (\ref{p3n})
we have
\begin{eqnarray}
X_\text{3rd}=\frac{P}{f^2}
\simeq x^2+2\frac{g}{f}xy\cos(\delta+\Delta),
\label{x3rd}
\end{eqnarray}
where the term $g^2y^2$ has been ignored again for simplicity.
Eq. (\ref{x3rd}) indicates that it is the second term in Eq. (\ref{x3rd})
that deviates the intercept $X_\text{3rd}$ of the red line
from the intercept $X_\text{JPARC}$ of the JPARC green line.
In order for the difference between $X_\text{JPARC}$ and
$X_\text{3rd}$ to be large, $f$ has to be small and
$|\cos(\delta+\Delta)|$ has to be large.
When $AL$ is small, in order for $f$ to be small,
$\left||\Delta m^2_{31}|L/4E-\pi\right|$ has to be small.
This is one of the conditions
to resolve the $\theta_{23}$ ambiguity.
Here we are using the reference value $\delta=\pi/4$,
so the deviation becomes maximal if $|\delta+\Delta|=|\pi/4+\Delta|\simeq\pi$.
In real experiments, however, nobody knows the value of the true $\delta$
in advance,
so it is difficult to design a long baseline experiment to
resolve the $\theta_{23}\leftrightarrow\pi/2-\theta_{23}$ ambiguity.
If $\delta$ turns out to satisfy $|\cos(\delta+\Delta)|\sim1$
in the result of the third experiment, then we may be able to resolve
the $\theta_{23}$ ambiguity as a byproduct.

\subsection{$\bar{\nu}_\mu\rightarrow\bar{\nu}_e$}
\begin{figure}
\vglue 1.5cm
\includegraphics[width=8cm]{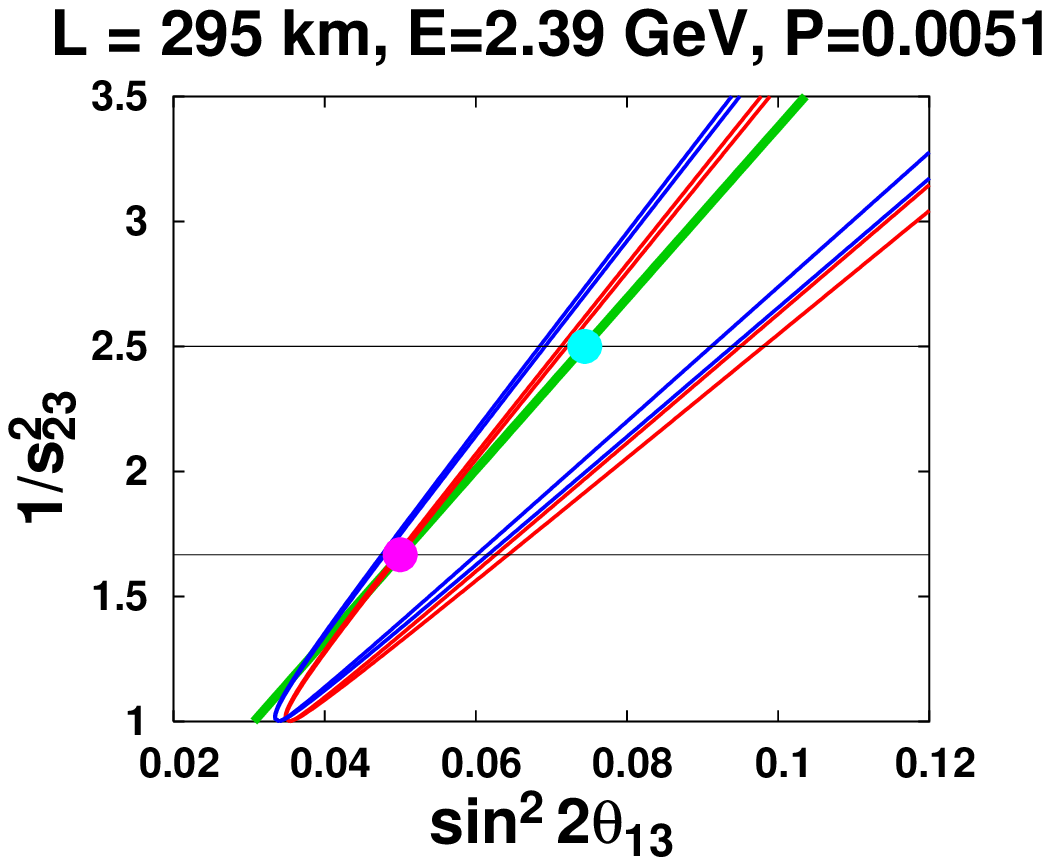}
\includegraphics[width=8cm]{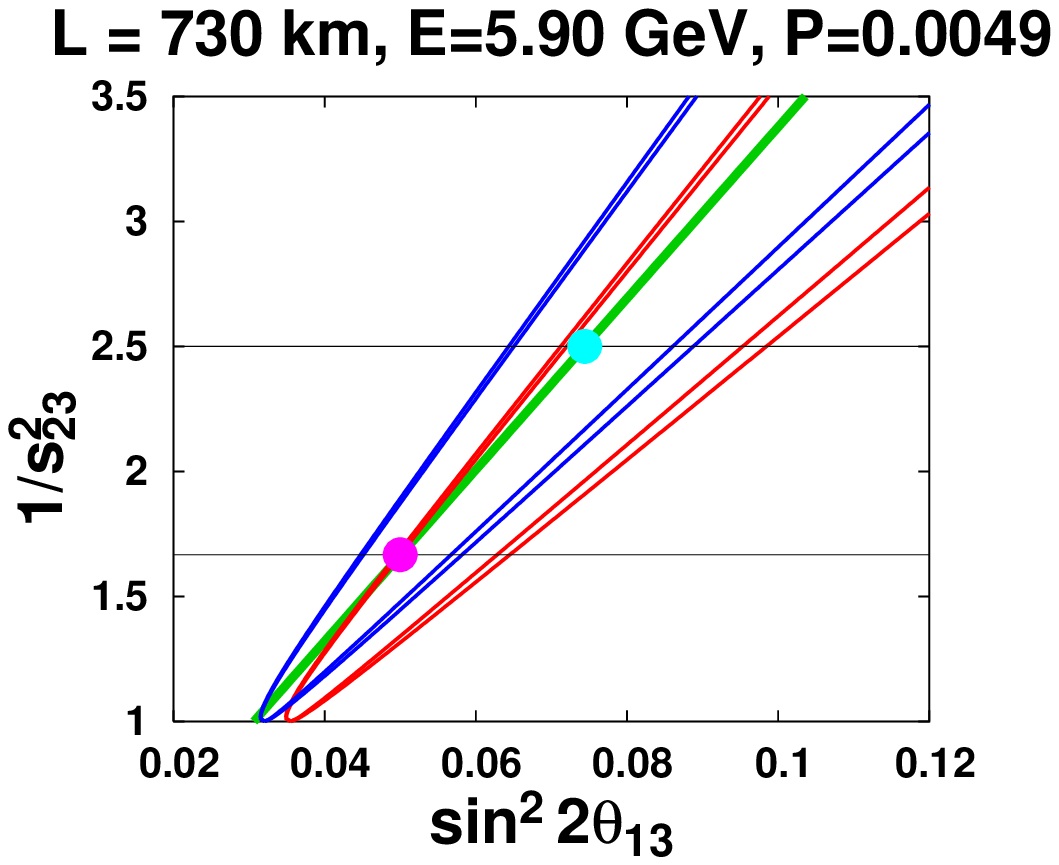}
\includegraphics[width=8cm]{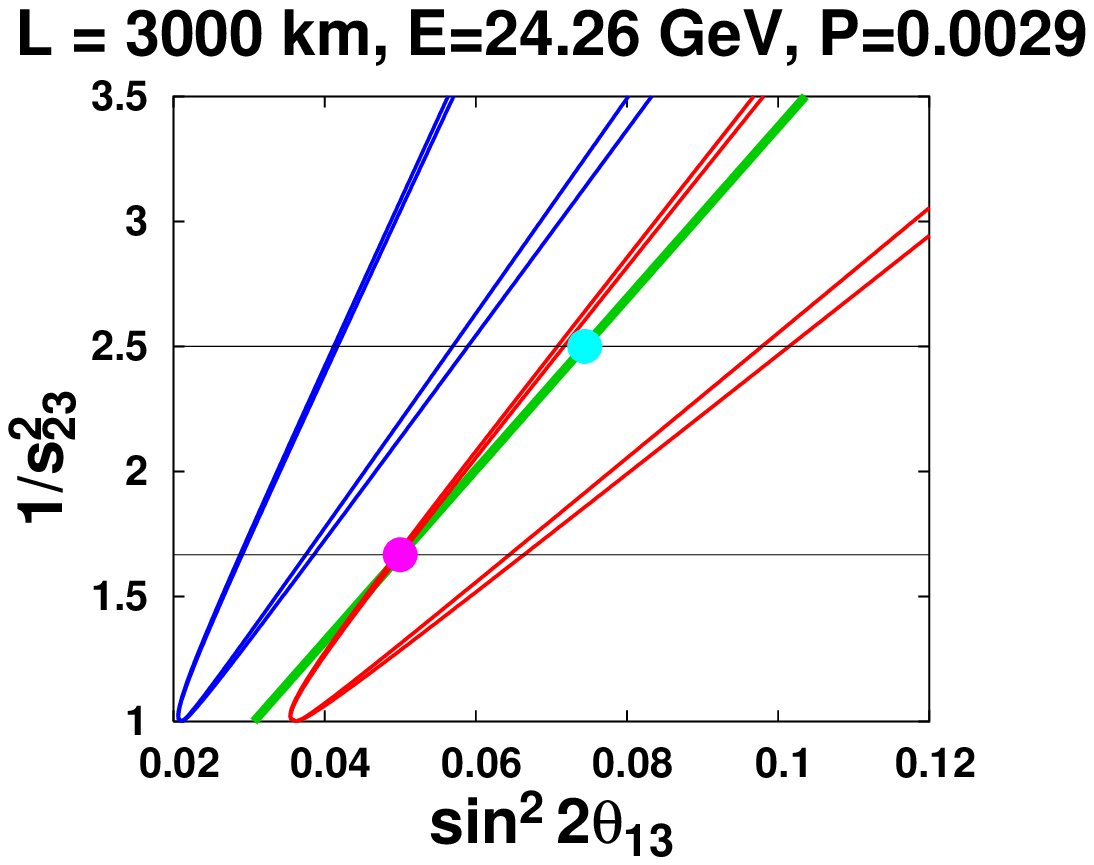}
\includegraphics[width=8cm]{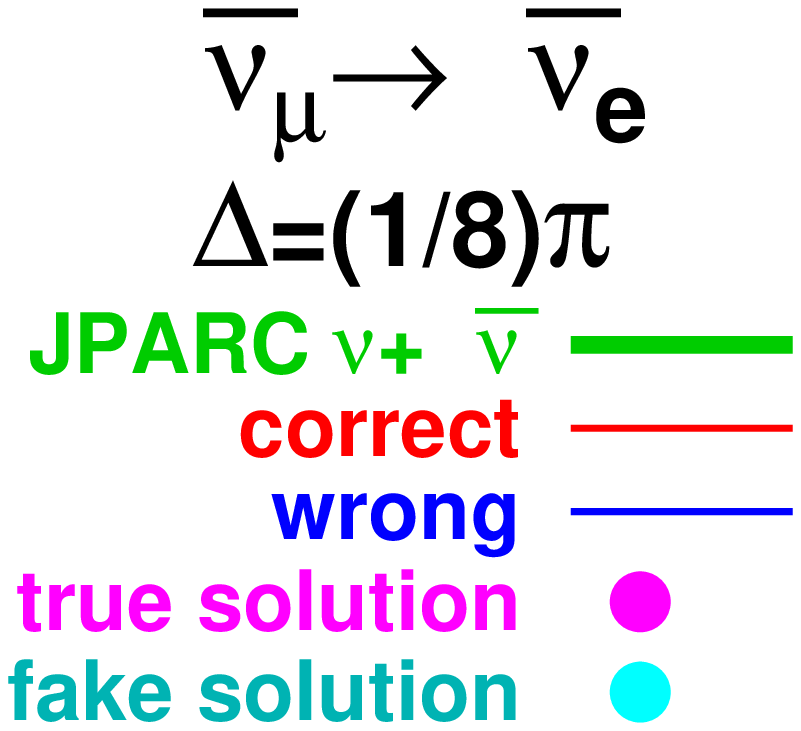}
\caption{\small
The trajectories of $P(\bar{\nu}_\mu\rightarrow\bar{\nu}_e)=$ const.
of the third experiment with
$\Delta\equiv|\Delta m^2_{31}|L/4E=\pi/8$ after JPARC.
The behaviors are almost similar to those for
$P(\nu_\mu\rightarrow\nu_e)=$ const.
The true values are those in Eq. (\ref{ref}).}
\label{fig10}
\vglue -0.5cm 
\end{figure}

It turns out that the situation does not change very much even if
we use the $\bar{\nu}_\mu \rightarrow \bar{\nu}_e$ channel
in the third experiment.
Typical curves are given for $\bar{\nu}_\mu \rightarrow \bar{\nu}_e$
in Fig.\,\ref{fig10}, which are similar to those in
Fig.\,\ref{fig7},\ref{fig8},\ref{fig9}.  Thus the conclusions
drawn on resolution of the ambiguities hold qualitatively
in the case of $\bar{\nu}_\mu \rightarrow \bar{\nu}_e$ channel.

\subsection{$\nu_e\rightarrow\nu_\tau$}
\begin{figure}
\vglue 1.5cm
\includegraphics[width=8cm]{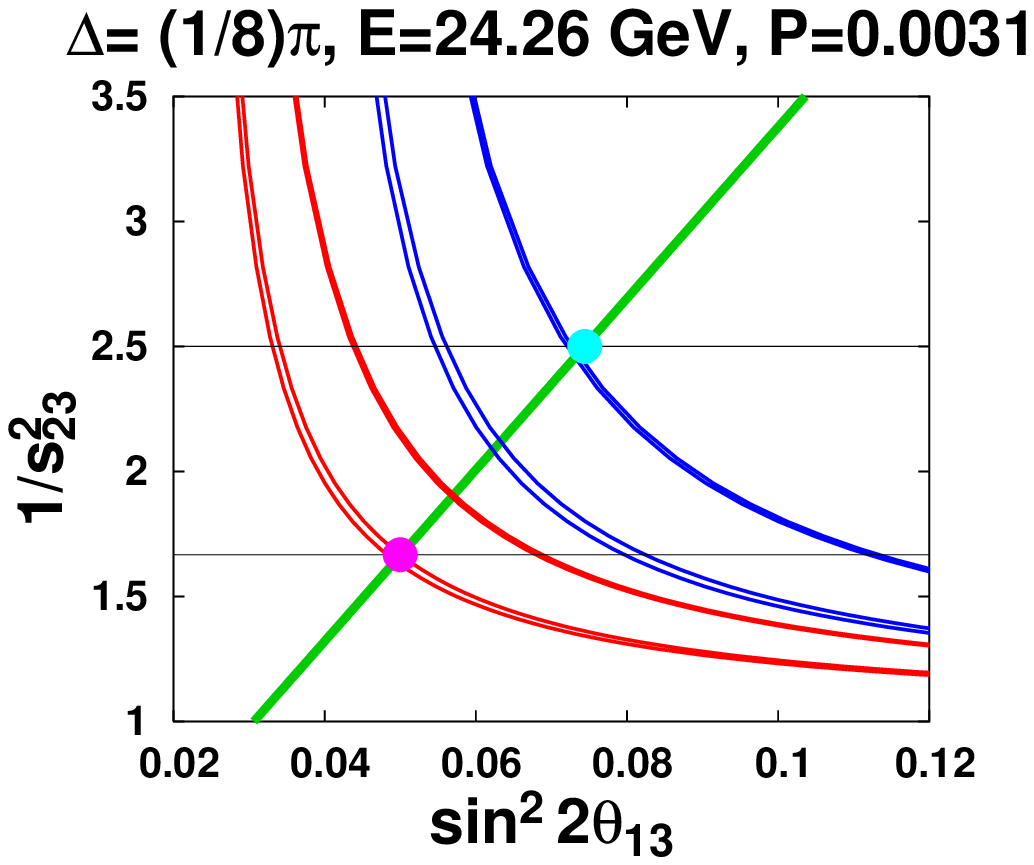}
\includegraphics[width=8cm]{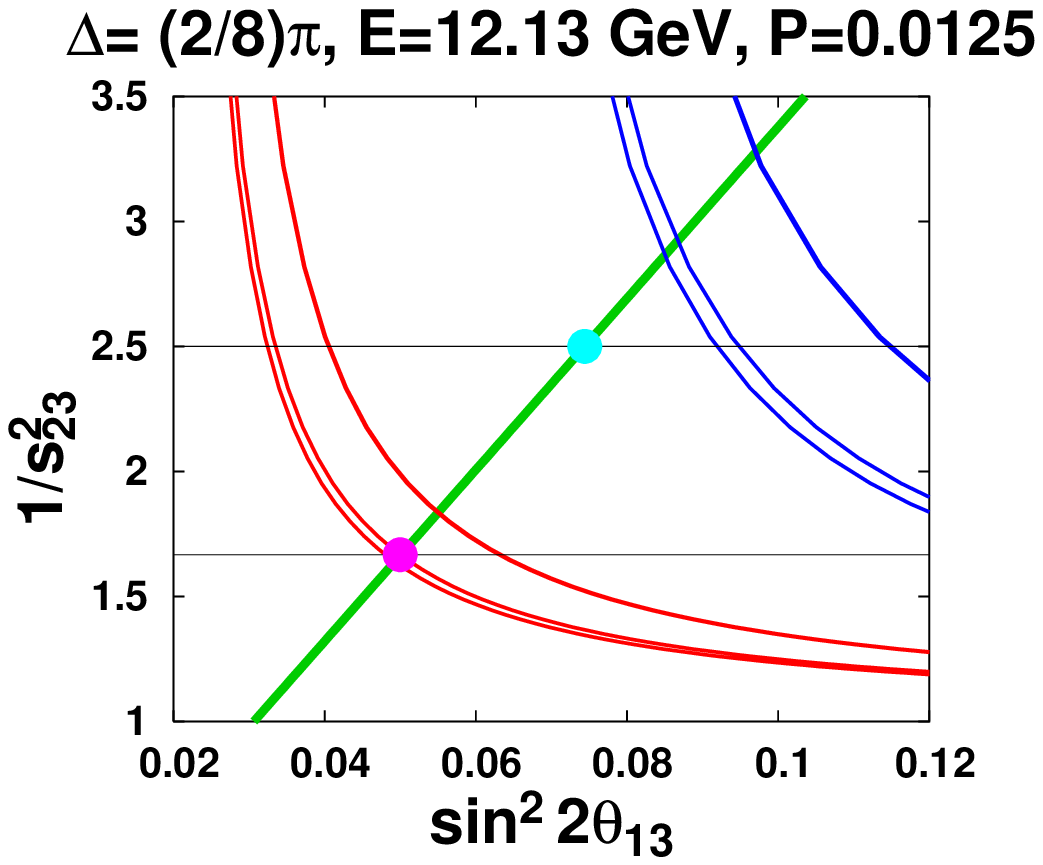}
\includegraphics[width=8cm]{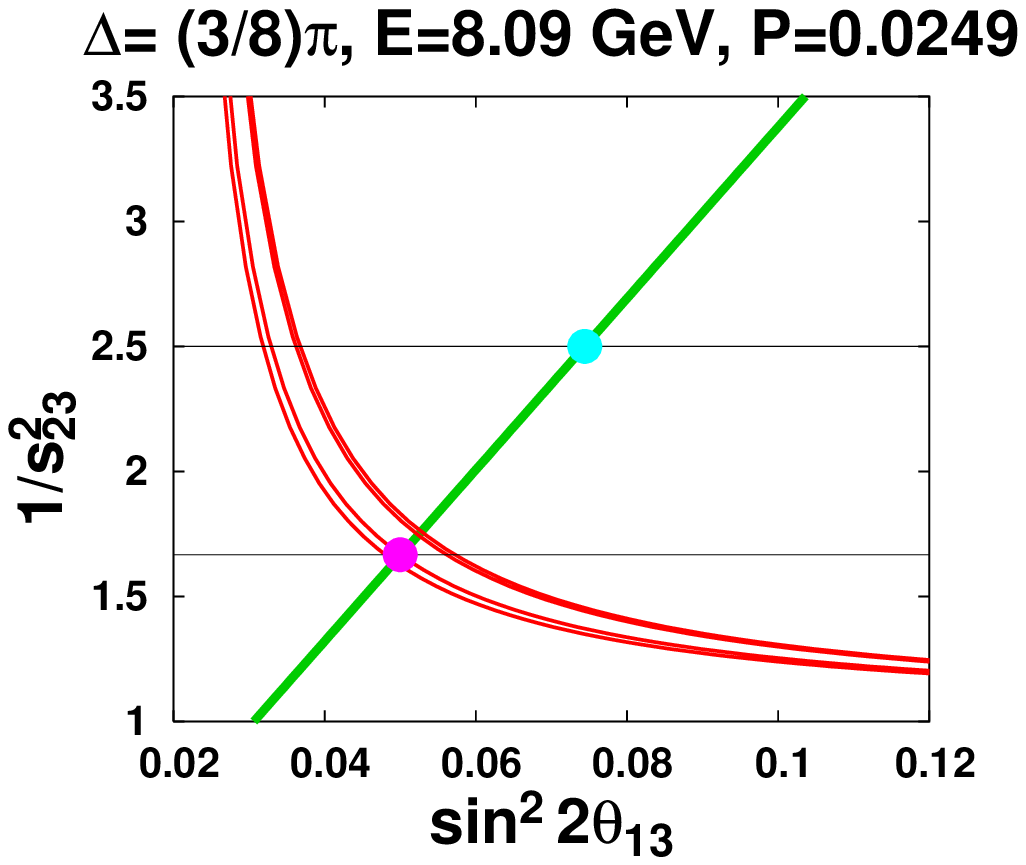}
\includegraphics[width=8cm]{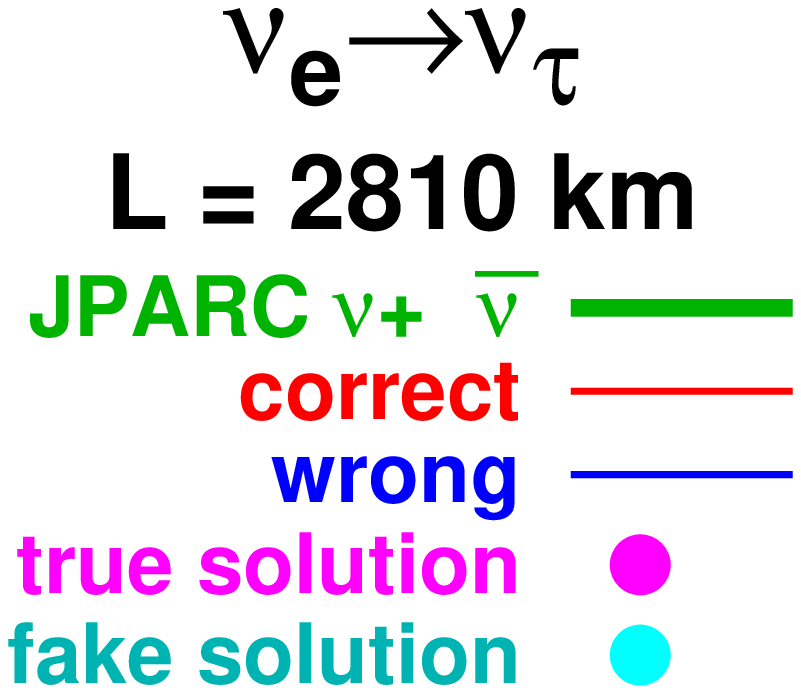}
\caption{\small
The trajectories of $P(\nu_e\rightarrow\nu_\tau)=$ const.
of the third experiment at $L$=2810km
with $\Delta\equiv|\Delta m^2_{31}|L/4E=(j/8)\pi~(j=1,2,3)$
after JPARC.  The true values are those in Eq. (\ref{ref}).
The curves intersect with the JPARC line perpendicularly,
so this channel is advantageous to resolve the ambiguities
from experimental point of view.
}
\label{fig11}
\vglue -0.5cm 
\end{figure}

The experiment with the channel $\nu_e\rightarrow\nu_\tau$
requires intense $\nu_e$ beams and it is expected that such
measurements can be done at neutrino factories or at
beta beam experiments \cite{Zucchelli:sa}.
The oscillation probability
$P(\nu_e\rightarrow\nu_\tau)$ is given by
\begin{eqnarray}
P(\nu_e\rightarrow\nu_\tau)
={\tilde x}^2f^2+2fg{\tilde x}{\tilde y}\cos(\delta+\Delta)
+{\tilde y}^2g^2,
\nonumber
\end{eqnarray}
where
\begin{eqnarray}
{\tilde x}&\equiv&c_{23}\sin2\theta_{23}\nonumber\\
{\tilde y}&\equiv&\left|\frac{\Delta m^2_{21}}{\Delta m^2_{31}}\right|
s_{23}\sin2\theta_{12},
\nonumber
\end{eqnarray}
and $f$, $g$ are given in Eqs. (\ref{f}) and (\ref{g}).
The solution for $P(\nu_e\rightarrow\nu_\tau)=Q$,
where $Q$ is constant, is given by
\begin{eqnarray}
X&=&\frac{Q}{f^2}\left\{\left[
1+\frac{2\cos^2(\delta+\Delta)}{1-C/Q}\right]
\frac{1-C/Q}{Y-1}+1\right.\nonumber\\
&{\ }&-\left.\frac{2\cos(\delta+\Delta)}{\sqrt{1-C/Q}}
\sqrt{\left[1+\frac{\cos^2(\delta+\Delta)}{1-C/Q}\right]
\frac{1-C/Q}{Y-1}+1}\right\},
\label{silver}
\end{eqnarray}
where $X\equiv\sin^22\theta_{13}$, $Y\equiv1/s^2_{23}$ as before
and $C$ is given in Eq. (\ref{c}).
Eqs. (\ref{silver}) is plotted in Fig.\,\ref{fig11} in the case of
$L$=2810km.
From Fig.\,\ref{fig11} we see that the curve
$P(\nu_e\rightarrow\nu_\tau)=Q$ intersects with the JPARC
green line almost perpendicularly, and it is experimentally
advantageous.  Namely, in real experiments all
the measured quantities have errors and the curves become
thick.  In this case the allowed region is small area
around the true solution in the $(\sin^22\theta_{13}, 1/s^2_{23})$
plane and one expects that the fake solution with respect to
the $\theta_{23}$ ambiguity can be excluded.
This is in contrast to the case of the $\nu_\mu\rightarrow\nu_e$
and $\bar{\nu}_\mu\rightarrow\bar{\nu}_e$ channels,
in which the slope of the red curves is almost the same
as that of the JPARC green line and the allowed region
can easily contain both the true and fake solutions,
so that it becomes difficult to distinguish the true point
from the fake one.

As in the case of the $\nu_\mu\rightarrow\nu_e$ channel,
the $\delta\leftrightarrow\pi-\delta$
ambiguity is expected to be resolved more likely
for the larger value of $|\Delta-\pi/2|$,
and the sgn($\Delta m^2_{31}$) ambiguity is resolved
easily for larger baseline $L$ (e.g., $L\sim$3000km).

Thus the measurement of the $\nu_e\rightarrow\nu_\tau$
channel is a promising possibility as a potentially powerful candidate
to resolve parameter degeneracies in the future.

\section{Discussion and Conclusion}
In this paper we have shown that the eightfold parameter degeneracy
in neutrino oscillations can be easily seen
by plotting the trajectory of constant probabilities
in the ($\sin^22\theta_{13}$, $1/s^2_{23}$) plane.
Using this plot, we have seen that the third measurement after
the JPARC results on $P(\nu_\mu\rightarrow\nu_e)$ and
$P(\bar{\nu}_\mu\rightarrow\bar{\nu}_e)$ may resolve
the sgn($\Delta m_{31}^2$) ambiguity at $L\gtrsim$1000km,
the $\delta\leftrightarrow\pi-\delta$ ambiguity off the
oscillation maximum
($\left|\Delta-\pi/2\right|\sim{\cal O}(1)$),
and the $\theta_{23}$ ambiguity if
$\left||\Delta m^2_{31}|L/4E-\pi\right|$ is small and
$\delta$ turns out to satisfy
$\left|\cos(\delta+\Delta)\right|\sim1$.
In general all these constraints on $\Delta\equiv|\Delta m^2_{31}|L/4E$
may be satisfied by
taking $\Delta=\pi$.  The condition $\Delta=\pi$, however,
actually corresponds to the oscillation minimum, and
the number of events is expected to be small for a number of reasons:
(1) The probability itself is small at the oscillation minimum;
(2) $\Delta=\pi$ implies low energy and the neutrino flux
decreases at low energy;
(3) The cross section is in general smaller at low energy
than that at high energy.
Therefore, to gain statistics, it is presumably wise to perform
an experiment at $\pi/2<\Delta<\pi$ after JPARC.
The off-axis NuMI experiment with 
$\pi/2<\Delta<\pi$ ($E\sim$1GeV)
may have advantage to resolve these ambiguities.

As is seen in Figs.\,\ref{fig8} and \ref{fig9},
the experiments at the oscillation maximum does not
appear to be useful after JPARC except for the sgn($\Delta m_{31}^2$)
ambiguity.
In order to achieve other goals such as resolution of
the $\delta\leftrightarrow\pi-\delta$ ambiguity and 
the $\theta_{23}$ ambiguity, it is wise to stay away from
$\Delta=\pi/2$ in experiments after JPARC.

Although only the oscillation probabilities were discussed without
taking the statistical and systematic errors into account in this
paper, we hope that the present work gives some insight
on how the ambiguities may be resolved in the future
long baseline experiments.

\appendix

\section{Expression for $P(\nu_\mu\rightarrow\nu_e)=P$\label{appendix1}}
First of all, let us derive Eqs. (\ref{p3n}) and (\ref{p3i}).
For the normal hierarchy, the probability $P(\nu_\mu \to \nu_e)=P$
is given by
\begin{eqnarray}
P&=&P(\nu_\mu \to \nu_e)
=x^2 f^2 + 2 x y f g
\cos\left(\delta+\Delta\right)
+ y^2 g^2\nonumber\\
&=&f^2\frac{X}{Y}+2f\sqrt{\frac{X}{Y}}\sqrt{C\left(1-\frac{1}{Y}\right)}
\cos\left(\delta+\Delta\right)+C\left(1-\frac{1}{Y}\right),
\label{a1}
\end{eqnarray}
where $X\equiv\sin^22\theta_{13}$, $Y\equiv 1/s^2_{23}$ as in the text,
$f$ is defined in Eq. (\ref{f}),
and $C$ is given in Eq. (\ref{c}).
Eq. (\ref{a1}) is rewritten as
\begin{eqnarray}
(P-C)(Y-1)+P-f^2X=2\sqrt{f^2X}\sqrt{C(Y-1)}
\cos\left(\delta+\Delta\right).
\label{a2}
\end{eqnarray}
Taking the square of the both hand sides of Eq. (\ref{a2}),
we get \footnote{Here we consider for simplicity the case where all the
arguments of the square root are positive.
After we obtain the final result, we see that the final formula
makes sense as long as the whole product of all the arguments is positive.}
\begin{eqnarray}
(f^2X)^2-2f^2X\left\{\left[P-C+2C
\cos^2\left(\delta+\Delta\right)\right](Y-1)+P\right\}
+\left[(P-C)(Y-1)+P\right]^2=0.
\nonumber
\end{eqnarray}
Solving this quadratic equation, we obtain
\begin{eqnarray}
\hspace*{-10mm}
f^2X&=&\left[P-C+2C
\cos^2\left(\delta+\Delta\right)\right](Y-1)+P\nonumber\\
&{\ }&\pm\sqrt{\left\{\left[P-C+2C
\cos^2\left(\delta+\Delta\right)\right](Y-1)+P\right\}
-\left[(P-C)(Y-1)+P\right]^2}\nonumber\\
&=&\left[P-C+2C\cos^2\left(\delta+\Delta\right)\right](Y-1)+P\nonumber\\
&{\ }&\pm2\cos\left(\delta+\Delta\right)\sqrt{C(Y-1)}
\sqrt{\left[P-C+C\cos^2\left(\delta+\Delta\right)\right](Y-1)+P}.
\label{a3}
\end{eqnarray}
If $\cos\left(\delta+\Delta\right)>0$ then from Eq. (\ref{a2})
we see that $(P+C)(Y-1)+P-f^2X$ has to be positive.
On the other hand, Eq. (\ref{a3}) gives
\begin{eqnarray}
&{\ }&(P+C)(Y-1)+P-f^2X\nonumber\\
&=&\mp2\cos\left(\delta+\Delta\right)\sqrt{C(Y-1)}\nonumber\\
&{\ }&\times
\left\{\sqrt{\left[P-C+C\cos^2\left(\delta+\Delta\right)\right](Y-1)+P}
-\cos\left(\delta+\Delta\right)\sqrt{C(Y-1)}\right\}.
\label{a4}
\end{eqnarray}
From Eq. (\ref{a4}) we conclude that we have to take the minus sign in
Eq. (\ref{a3}) for the right hand side of Eq. (\ref{a4}) to be positive.
Hence from $P(\nu_\mu \to \nu_e)=P$ we get
\begin{eqnarray}
f^2X&=&\left[P-C+2C\cos^2(\delta+\Delta)\right](Y-1)+P\nonumber\\
&{\ }&-2\cos(\delta+\Delta)\sqrt{C(Y-1)}
\sqrt{\left[P-C\sin^2(\delta+\Delta)\right](Y-1)+P},
\label{a5}
\end{eqnarray}
and from $P(\bar{\nu}_e \to \bar{\nu}_\mu)=\bar{P}$ we have
\begin{eqnarray}
\bar{f}^2X&=&\left[\bar{P}-C+2C\cos^2(\delta-\Delta)\right](Y-1)+\bar{P}\nonumber\\
&{\ }&-2\cos(\delta-\Delta)\sqrt{C(Y-1)}
\sqrt{\left[\bar{P}-C\sin^2(\delta-\Delta)\right](Y-1)+\bar{P}}.
\label{a6}
\end{eqnarray}
When $P-C\sin^2(\delta+\Delta)>0$, Eq. (\ref{a5})
is a hyperbola, and the physical region for $Y-1$ is $Y-1\ge0$.
On the other hand, when $P-C\sin^2(\delta+\Delta)<0$,
Eq. (\ref{a5}) becomes an ellipse and the physical region for $Y-1$ is
$0\le Y-1\le P/[C\sin^2(\delta+\Delta)-P]$.

Similarly, we obtain for the inverted hierarchy:
\begin{eqnarray}
\bar{f}^2X&=&\left[P-C+2C\cos^2(\delta-\Delta)\right](Y-1)+P\nonumber\\
&{\ }&+2\cos(\delta-\Delta)\sqrt{C(Y-1)}
\sqrt{\left[P-C\sin^2(\delta-\Delta)\right](Y-1)+P},
\label{a7}
\end{eqnarray}
\begin{eqnarray}
f^2X&=&\left[\bar{P}-C+2C\cos^2(\delta+\Delta)\right](Y-1)+\bar{P}\nonumber\\
&{\ }&+2\cos(\delta+\Delta)\sqrt{C(Y-1)}
\sqrt{\left[\bar{P}-C\sin^2(\delta+\Delta)\right](Y-1)+\bar{P}}.
\label{a8}
\end{eqnarray}

\section{Trajectories at the oscillation maximum\label{appendix2}}
\begin{figure}
\vglue 3cm 
\includegraphics[scale=1.0]{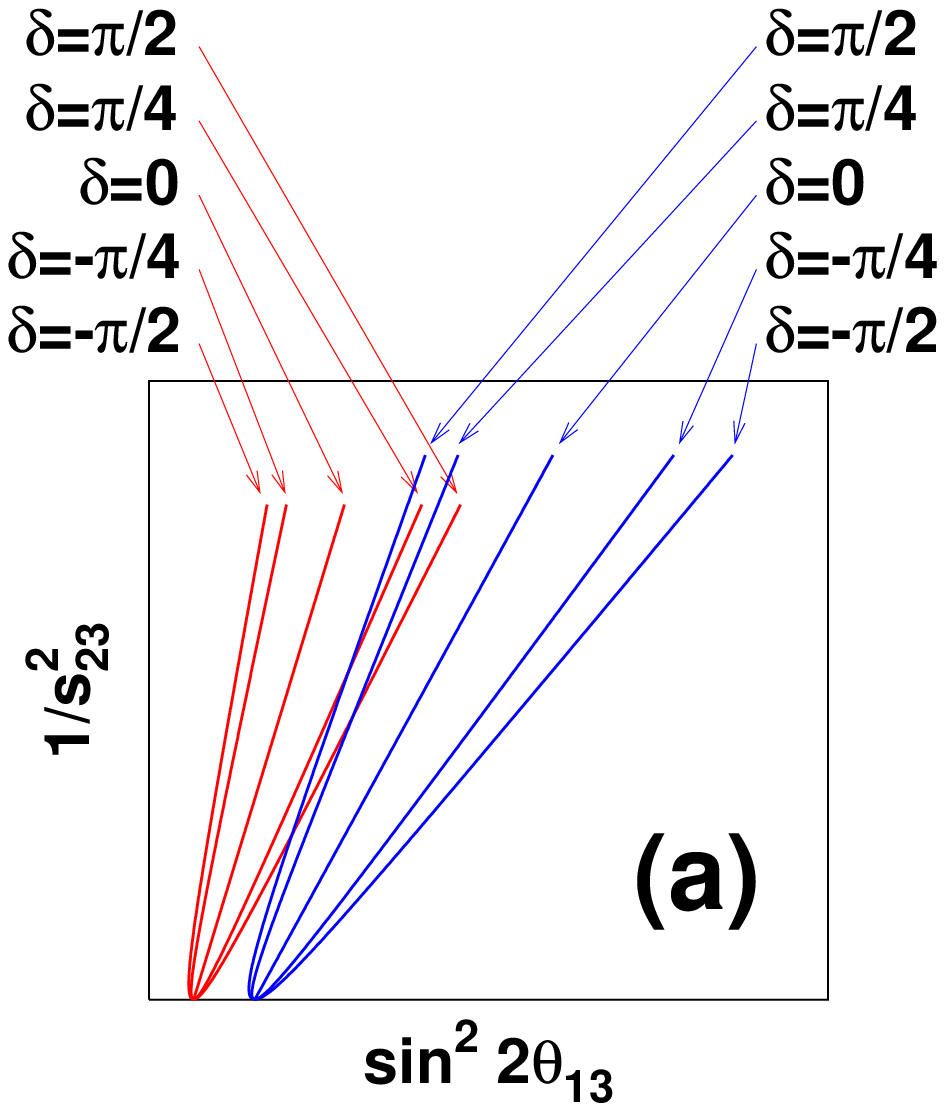}
\includegraphics[scale=1.0]{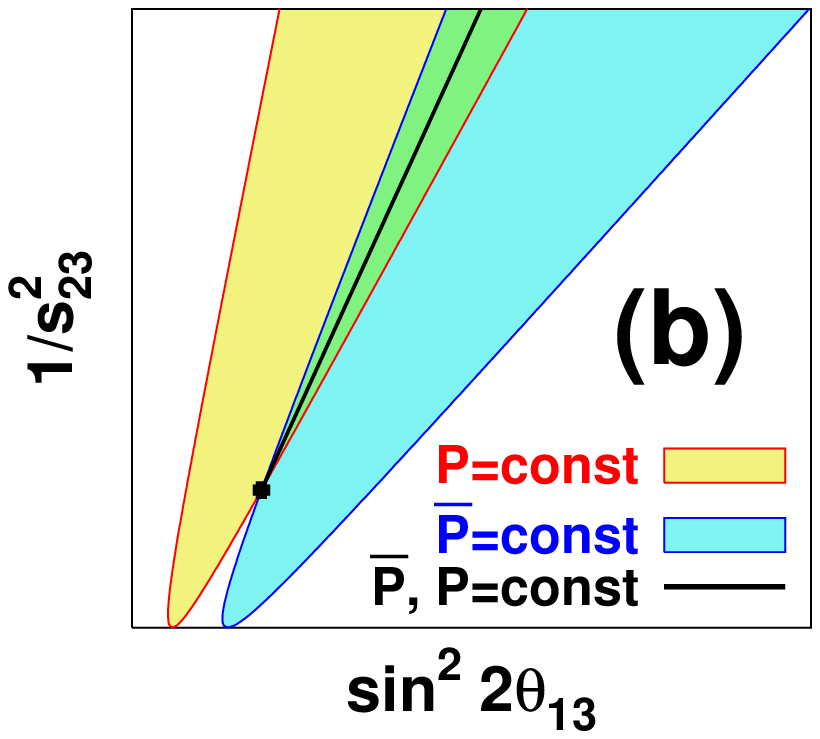}
\vglue 0.5cm
\caption{The region of constant probabilities at the
oscillation maximum. (a) Each red (blue) line stands for
$P(\nu_\mu\rightarrow\nu_e)=$ const.
($P(\bar{\nu}_\mu\rightarrow\bar{\nu}_e)=$ const.) for a specific
value of $\delta$.  The red line on the right (left) edge corresponds to
$\delta=+\pi/2$ ($\delta=-\pi/2$), while the blue line on the edge right (left)
corresponds to $\delta=-\pi/2$ ($\delta=+\pi/2$).
(b) When $\delta$ varies from $0$ to $2\pi$,
the line $P(\nu_\mu\rightarrow\nu_e)=$ const. sweeps out
the yellow region, whereas the line $\bar{\nu}_\mu\rightarrow\bar{\nu}_e)=$ const.
sweeps out the light blue region.  The black straight line,
which is given by $P(\nu_\mu\rightarrow\nu_e)=$ const. {\it and}
$P(\bar{\nu}_\mu\rightarrow\bar{\nu}_e)=$ const., lies in the overlapping
green region.}
\label{fig12}
\end{figure}

Throughout this appendix we will assume $\Delta=\pi/2$ and we will
assume $\Delta m^2_{31}>0$ for most part of this appendix.
From Eq. (\ref{a5}), the condition $P(\nu_\mu \to \nu_e)=P$ for the
neutrino mode alone gives
\begin{eqnarray}
f^2X&=&\left[P-C+2C\sin^2\delta\right](Y-1)+P\nonumber\\
&{\ }&+2\sin\delta\sqrt{C(Y-1)}
\sqrt{\left[P-C\cos^2\delta\right](Y-1)+P},
\label{b1}
\end{eqnarray}
while the condition $P(\bar{\nu}_\mu \to \bar{\nu}_e)=\bar{P}$ for
the anti-neutrino mode alone gives 
\begin{eqnarray}
\bar{f}^2X&=&\left[\bar{P}-C+2C\sin^2\delta\right](Y-1)+\bar{P}\nonumber\\
&{\ }&-2\sin\delta\sqrt{C(Y-1)}
\sqrt{\left[\bar{P}-C\cos^2\delta\right](Y-1)+\bar{P}}.
\label{b2}
\end{eqnarray}
When $\delta$ ranges from $-\pi/2$ to $\pi/2$,
Eq. (\ref{b1}) sweeps out the inside of a hyperbola,
as is depicted by the red curves in Fig.\,\ref{fig12}(a), while
(\ref{b2}) sweeps out the inside of another hyperbola
for the anti-neutrino mode (cf. the blue curves in Fig.\,\ref{fig12}(a)).
Notice that the left (right) edge of the hyperbola (\ref{b1}) for the
neutrino mode corresponds to $\delta=-\pi/2$ ($\delta=+\pi/2$)
whereas the left (right) edge of the other hyperbola (\ref{b2}) for the
anti-neutrino mode corresponds to $\delta=+\pi/2$ ($\delta=-\pi/2$).
Since the straight line (\ref{omn}) is the intersection of the
two regions (the yellow and light blue regions in
Fig.\,\ref{fig12}(b)), the lowest point
in the straight line is obtained by putting $\delta=+\pi/2$
($\delta=-\pi/2$) if $P/f^2<\bar{P}/\bar{f}^2$ (if $P/f^2>\bar{P}/\bar{f}^2$),
respectively, depending on whether the region for the anti-neutrino mode
is to the right of that for the neutrino mode.
Therefore, if $P/f^2<\bar{P}/\bar{f}^2$, then putting $\delta=+\pi/2$
in Eqs. (\ref{om1}) and (\ref{om2})
and assuming $xf>yg$, which should hold if $\sin^22\theta_{13}$ is not so small,
we get
\begin{eqnarray}
\sqrt{P}=xf-yg=f\sqrt{\frac{X}{Y}}-\sqrt{C\left(1-\frac{1}{Y}\right)}\nonumber\\
\sqrt{\bar{P}}=x\bar{f}+yg=\bar{f}\sqrt{\frac{X}{Y}}+\sqrt{C\left(1-\frac{1}{Y}\right)}
\nonumber
\end{eqnarray}
which lead to the minimum value of $Y$
\begin{eqnarray}
Y^{\text{(n)}}_\text{min}
=\left[1-\frac{(f\sqrt{\bar{P}}-\bar{f}\sqrt{P})^2}{C(f+\bar{f})^2}
\right]^{-1}
\nonumber
\end{eqnarray}
for the normal hierarchy.  On the other hand, for the inverted hierarchy,
the corresponding values of $\delta$ for the edges for the two modes
are the same as those for the normal hierarchy ($\delta=\pm\pi/2$).
Hence, if $P/\bar{f}^2<\bar{P}/f^2$, then putting $\delta=+\pi/2$
in Eqs. (\ref{om3}) and (\ref{om4}) and assuming $x\bar{f}>yg$, we obtain
\begin{eqnarray}
\sqrt{P}=x\bar{f}-yg\nonumber\\
\sqrt{\bar{P}}=xf+yg
\nonumber
\end{eqnarray}
which leads to the minimum value of $Y$
\begin{eqnarray}
Y^{\text{(i)}}_\text{min}
=\left[1-\frac{(\bar{f}\sqrt{\bar{P}}-f\sqrt{P})^2}{C(f+\bar{f})^2}
\right]^{-1}
\nonumber
\end{eqnarray}
for $\Delta m^2_{31}<0$.

\begin{acknowledgments}
I would like to thank Hiroaki Sugiyama for many discussions.
This work was supported in part by Grants-in-Aid for Scientific Research
No.\ 16540260 and No.\ 16340078, Japan Ministry
of Education, Culture, Sports, Science, and Technology.
\end{acknowledgments}


\newpage
\pagestyle{empty}

\newpage

\newpage

\newpage

\newpage

\newpage

\newpage

\newpage

\newpage

\newpage

\newpage

\newpage

\end{document}